\documentclass[fleqn,usenatbib]{mnras}

\usepackage{newtxtext,newtxmath}

\usepackage[T1]{fontenc}
\usepackage{ae,aecompl}

\usepackage{graphicx}	
\usepackage{amsmath}	

\DeclareMathOperator{\dtd}{DTD}

\title[Density of the MW's corona at $z=1.6$]{The density of the Milky Way's corona at $z\approx1.6$ through ram pressure stripping of the Draco dSph galaxy}

\author[A.E. Gr\o nnow et al.]{
Asger Gr\o nnow,$^{1}$
Filippo Fraternali,$^{1}$\thanks{E-mail: fraternali@astro.rug.nl}
Federico Marinacci,$^{2}$
Gabriele Pezzulli,$^{1}$\newauthor
Eline Tolstoy,$^{1}$
Amina Helmi,$^{1}$
and Anthony G.A. Brown$^{3}$
\\
$^{1}$Kapteyn Astronomical Institute, University of Groningen, 9700AV Groningen, The Netherlands\\
$^{2}$Department of Physics \& Astronomy, University of Bologna, via Gobetti 93/2, 40129 Bologna, Italy\\
$^{3}$Leiden Observatory, Leiden University, Niels Bohrweg 2, 2333 CA Leiden, The Netherlands
}

\date{Accepted 2023 December 19. Received 2023 November 06; in original form 2022 July 30}

\pubyear{2024}

\begin{document}
\label{firstpage}
\pagerange{\pageref{firstpage}--\pageref{lastpage}}
\maketitle

\begin{abstract}
Satellite galaxies within the Milky Way’s (MW) virial radius $R_{\mathrm{vir}}$ are typically devoid of cold gas due to ram pressure stripping by the MW’s corona. The density of this corona is poorly constrained today and essentially unconstrained in the past, but can be estimated using ram pressure stripping. In this paper, we probe the MW corona at $z\approx 1.6$ using the Draco dwarf spheroidal galaxy. We assume that i) Draco’s orbit is determined by its interaction with the MW, whose dark matter halo we evolve in time following cosmologically-motivated prescriptions, ii) Draco’s star formation was quenched by ram pressure stripping and iii) the MW’s corona is approximately smooth, spherical and in hydrostatic equilibrium. We used GAIA proper motions to set the initial conditions and Draco’s star formation history to estimate its past gas content. We found indications that Draco was stripped of its gas during the first pericentric passage. Using 3D hydrodynamical simulations at a resolution that enables us to resolve individual supernovae and assuming no tidal stripping, which we estimate to be a minor effect, we find a density of the MW corona $\geq 8\times 10^{-4}$ cm$^{-3}$ at a radius $\approx 0.72R_{\mathrm{vir}}$. This provides evidence that the MW’s corona was already in place at $z\approx 1.6$ and with a higher density than today. If isothermal, this corona would have contained all the baryons expected by the cosmological baryon fraction. Extrapolating to today shows good agreement with literature constraints if feedback has removed $\lesssim 30$\% of baryons accreted onto the halo.
\end{abstract}




\section{Introduction}
Relatively massive galaxies such as our Milky Way (MW) are expected from theoretical galaxy formation and cosmological simulations to contain a hot gas corona extending to roughly the virial radius $R_{\mathrm{vir}}$. This corona is formed by gas that is shock heated as it falls into the dark matter (DM) halo and cannot cool efficiently \citep{rees77,white91}. Instead, it settles into a hydrostatic atmosphere between the interstellar medium (ISM) and the intergalactic medium (IGM) or local group medium. The MW corona is expected to have formed roughly at redshift $z\approx 2$ when the MW reached a virial mass of a few $\times 10^{11} M_\odot$ \citep{keres09,correa18}. In the MW the existence of this hot coronal gas has been established relatively close to the galactic disc through observations in absorption and X-ray emission \citep[e.g.][]{gupta12,henley13,miller13,li17,bregman18}. The head-tail morphology of many High-Velocity Clouds also shows that they are interacting with a surrounding medium \citep{putman11} but these clouds are generally also found to be at distances of $d\lesssim 10$ kpc \citep{thom08,lehner22}. These thus provide indirect evidence for the presence of the corona in the vicinity of the disc.

However, observations of the H\textsc{i} gas content of the MW satellite galaxies support the expectation that the corona extends much further out to roughly the MW's virial radius of about 250 kpc \citep{putman21}. Generally, satellites within $R_{\mathrm{vir}}$ are found to have no detectable H\textsc{i} down to very low upper limits, while satellites outside of $R_{\mathrm{vir}}$ have substantial H\textsc{i} masses. This suggests that satellites lose their gas due to ram pressure stripping by the hot corona. The notable exceptions are the Magellanic Clouds which are relatively massive and probably on their first infall \citep{besla07}. Tidal stripping can also remove gas from nearby satellites but this is not expected to be a major effect in most cases \citep{gatto13,putman21}.

This ram pressure stripping can be used not only as evidence of the existence of the hot corona but also to put constraints on its density along part of a satellite's orbit \citep{grcevich09,gatto13,salem15,putman21}. This is because the ram pressure is given by the density and relative velocity of the surrounding medium along the orbit $P_{\mathrm{ram}}=\rho_{\mathrm{cor}} v^2$ \citep{Gunn72}. Assuming that the satellite is spherical and that the stripping occurs instantly at pericentre, where the velocity is greatest, a lower limit on the particle density of the corona at pericentre can be roughly estimated from
\begin{equation}
\label{eq:ramstrip}
n_{\mathrm{cor}} \gtrsim \frac{\sigma_\star^2 n_{\mathrm{ISM}}}{v_{\mathrm{peri}}^2},
\end{equation}
where $\sigma_\star$ is the stellar velocity dispersion of the satellite, $n_{\mathrm{ISM}}$ is the particle density of the satellite's ISM, and $v_{\mathrm{peri}}$ is the velocity with respect to the corona at pericentre \citep{mori00,grcevich09}. However, this is only a rough estimate and the assumption that all the stripping occurs at pericentre is often not valid \citep{gatto13}.

In simulations, the stripping throughout the orbit can be included, as well as other important effects on the gas such as supernova feedback. Many such simulations of satellite galaxies undergoing ram pressure stripping have been examined in the literature \citep[e.g.][]{mayer06,gatto13,salem15,nichols15,emerick16,tepper-garcia18,hausammann19,tepper-garcia19}. Most of these works do not attempt to constrain the MW coronal density but rather adopt a single value or profile and focus on other aspects. The exceptions are \cite{gatto13} and \cite{salem15} who each ran an array of simulations with different coronal densities in order to constrain the average density around the satellites. \cite{salem15} simulated the partial ram pressure stripping of the LMC during its recent pericentric passage to constrain the present day coronal density at its distance of $r\approx 50$ kpc from the centre of the MW. \cite{gatto13} used a more general approach based on the observed star formation history (SFH) to simulate the stripping of the Carina and Sextans dwarf spheriodal galaxies. The SFH together with the calculated orbit reveals which passage stripped the last of the dwarf's ISM. By assuming a spherical isothermal density profile the density of the dwarf's ISM before this passage can then be derived from its star formation rate (SFR). In this way, they found lower limits in the coronal density at distances of $40-90$ kpc from the Galactic centre.
In the analysis of \cite{gatto13} the main uncertainty was in the observed proper motions of the satellites, which greatly affect the derived orbits. These proper motion measurements have improved dramatically in recent years with the advent of \emph{Gaia} \citep{gaia16}. A further limitation was the two dimensional geometry assumed in their simulations.

The density of the corona is particularly important for the outstanding question of the so-called \emph{missing baryons}. On the cosmological scale this refers to the problem that censuses of observed baryons in the Universe generally fall significantly short of the cosmological baryon fraction \citep{shull12,nicastro17}. The cosmological baryon fraction, defined as the ratio between the total baryonic and DM masses, is known to good accuracy from cosmology to be $f_\mathrm{b}=\Omega_b/\Omega_c \approx 0.18$ \citep{planck20}, where $\Omega_b$ and $\Omega_c$ are the density parameters for baryons and DM assuming a $\Lambda$ cold dark matter cosmology. These baryons are expected to mainly reside in a warm-hot intergalactic medium. The recent analyses of \cite{nicastro18,macquart20} can account for all of the baryons although with large uncertainties. However, there is also a missing baryon problem on the scale of individual galaxies. In the MW the stellar and cold gas mass is $\approx 6\times 10^{10} M_\odot$ \citep{bland-hawthorn16} while the virial mass is $\approx 10^{12} M_\odot$ \citep{callingham19,posti19}. Hence, the MW would need a hot corona of $\approx 10^{11} M_\odot$ to account for all the baryons expected within its DM halo from the cosmological baryon fraction. However, the corona could be less massive than this if feedback has expelled gas from the halo and/or prevented gas from falling within the virial radius to begin with. Observational studies are inconclusive on this with some estimating that the corona only contains a small fraction of the baryonic mass \citep{anderson10,li17,bregman18} and others that it could contain all the missing baryons \citep{gupta12,faerman17,martynenko22}. In any case, the mass of the corona should not exceed the mass calculated from the cosmological baryon fraction. With an assumed density profile this can be used to put upper limits on its density \citep[e.g.][]{tepper-garcia15}.

In this paper we estimate the density of the corona using the Draco dwarf spheroidal galaxy. This satellite is an ideal target to probe the MW corona through its ram pressure stripping. It does not show signs of tidal interaction \citep{segall07,munoz18} and contains no detectable H\textsc{i} gas down to a very low upper limit \citep{putman21}. Its SFH suggests that it lost its gas already around 10 Gyr ago (see Section \ref{sec:sfh}). This is too late to be explained by reionization but after the corona is expected to have formed. Using the highly accurate proper motions from \emph{Gaia} Early Data Release 3 \citep{li21} we find that the drop in Draco's star formation aligns well with its first infall. We take advantage of this to constrain the density of the MW's early corona, for the first time, by simulating Draco's first passage. 
Our final results rely on a few assumptions. 
We assume that Draco's orbit is largely determined, at least since $z\approx 1.6$, by the potential of the Milky Way's dark matter halo, which we evolve in time following prescriptions from cosmological simulations, without significant effects due to interactions with other structures. 
Such an assumption is clearly valid for the main possible perturbers, i.e.\ the Magellanic Clouds as they are and have always been located in the opposite hemisphere with respect to Draco \citep{Fritz+2018,Patel+2020}.
We also assume that Draco's star formation was quenched by ram pressure stripping against the Milky Way's corona and not by an encounter with another satellite, which is expected to be a rare event \citep{Genina+2022}.
Finally, we assume that the Milky Way's corona is smooth, roughly spherical and in hydrostatic equilibrium at least to first order since $z\approx 1.6$ to now. 
Our assumptions and their implications are discussed further in the latter half of Section \ref{sec:discussion}.

The paper is structured as follows. In Section \ref{sec:methods} we describe Draco's SFH, integrate its orbit, and describe our simulation setup. In Section \ref{sec:results} we show and discuss our density constraint on the early corona and how it relates to the present day literature constraints through accretion and outflows. In Section \ref{sec:discussion} we describe how our results depend on the assumed virial mass, coronal temperature, and the resolution of the simulations, and discuss the limitations of our method. Finally, we summarize and conclude in Section \ref{sec:conclusions}. A simple model for the growth of the mass in the corona, which we use to extrapolate some of our results, is described in Appendix \ref{sec:coronaevol}.


\section{Methods}
\label{sec:methods}
We follow the ram pressure stripping of Draco by the MW corona by simulating a volume containing Draco's cold gas and gravitational potential and part of the surrounding lower density coronal gas. We simulate Draco travelling through the corona by injecting a `wind' with varying velocity into this simulation volume from one of the boundaries. In doing this we neglect the changing direction of the trajectory throughout the orbit which is unimportant for our purposes that are only concerned with the efficiency of ram pressure stripping. While we will generally assume a constant coronal density along the orbit (but see Section \ref{sec:vardens}), the change in velocity is important due to the fact that ram pressure is strongly velocity dependent $P_\mathrm{ram}\propto v^2$. Hence, before running the simulations we need to integrate Draco's orbit back in time from its present day position and velocity in order to find this time dependent velocity to be used in the simulations. In addition, the simulated orbits, together with Draco's star formation history (SFH), have to be consistent with its ISM having been lost to ram pressure stripping during the first passage. We describe this SFH and our orbit calculations, as well as the numerical setup for our simulations of Draco, in the following sections.

\subsection{Star formation history}
\label{sec:sfh}
\begin{figure}
    \centering
    \includegraphics[width=0.49\textwidth]{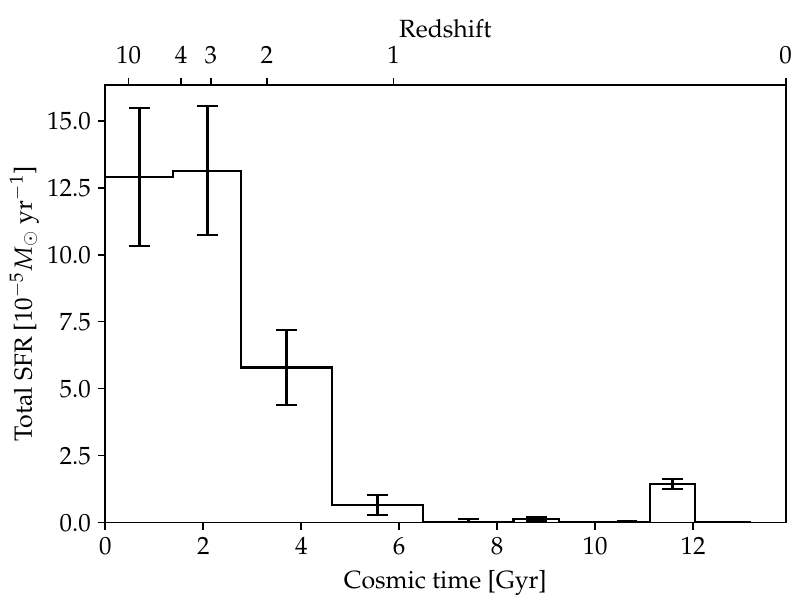}
    \caption{Total SFR of Draco as function of time from \protect\cite{aparicio01}. This is the sum of the SFRs that they derive for radii $r<7.5'$ and $7.5'<r<30'$ with the uncertainties added in quadrature. The time axis has been slightly rescaled from a present day cosmic time of 15 Gyr to 13.88 Gyr as described in the text.}
    \label{fig:sfh}
\end{figure}

We use the SFH of Draco from \cite{aparicio01} for evaluating the orbits as well as for the initial conditions and SNIa feedback rates (while SNII rates are instead calculated from the local SFR in each cell, see Section \ref{sec:feedback}), as described later in this section. This SFH begins 15 Gyr ago so we rescale the time of the bins slightly such that it begins at 13.88 Gyr ago to be consistent with our assumed cosmology for the evolution of the MW potential (see Section \ref{sec:orbitint}). This 8 per cent rescaling of the time is negligible compared to the width of the bins. \cite{aparicio01} reports the SFH within an inner region of $r<7.5'$ and an outer region of $7.5'<r<30'$ which we sum to get the total SFR within $r<30'$ which is $\approx 700$ pc. We show this SFH in Figure \ref{fig:sfh}. Note that the small bump around 2-3 Gyr ago is likely not actual star formation but rather caused by `blue straggler' stars \citep{mapelli07,munoz18}. From the bin centred at cosmic age $\approx 3.5$ Gyr to the next bin at $\approx 5.5$ Gyr the SFR drops from being clearly above zero to consistent with zero (within 1.2$\sigma$), which implies that Draco lost its gas during this time. This drop is consistent with the other published SFHs of Draco in \cite{dolphin03,weisz14}. There is also a drop from the earlier bin at a cosmic age around 2 Gyr. However, it is not clear if the corona had already formed by then and in any case we are not aiming to reproduce the entire SFH but only the last stripping event. Also, as we show in the following section, for reasonable MW masses Draco's first infall occurs later as well. Hence, we will focus only on the last big drop in the SFH. The earlier decrease might have been caused by feedback and/or stripping from gas outside the virial radius. While we show in Section \ref{sec:effectoffb} that feedback alone is not efficient at removing gas, it still tends to lower the SFR by spreading out the gas. Additionally, \cite{putman21} recently found that the observed gas masses and positions of MW and M31 satellites suggest that ram pressure stripping from a local group medium surrounding both galaxies can be effective outside the virial radius of either galaxy.

\subsection{Orbit integration}
\label{sec:orbitint}
We use the distance and velocity of Draco from the \emph{Gaia} Early Data Release 3 (EDR3) reported in \cite{li21} as the starting point for integrating Draco's orbit back in time. These are $d_{\mathrm{GC}}=81.8_{-5.7}^{+6.1}$ kpc for the distance from the Galactic centre and $v_{\mathrm{tot}}=181.5_{-7.5}^{+7.2}$ km s$^{-1}$ for the total velocity in the Galactic rest frame. We take into account the evolution of the MW potential throughout the orbit according to the change in virial mass, radius, and concentration of its Navaro-Frank-White \citep[NFW,][]{navarro96} DM profile from the halo evolution models of \cite{zhao09}. We calculate these models using the code made available by these authors\footnote{This code is available at \url{http://www.shao.ac.cn/dhzhao/mandc.html}} with updated cosmological parameters from \cite{planck20}. We do not include any separate components for baryonic matter because the DM dominates the potential at all times and because the evolution of the baryonic components are much more uncertain. Hence, the baryonic contribution is, in principle, included in this virial mass. However, within the uncertainty in the mass of the MW the distinction between DM halo mass and total DM + baryonic mass is not important. 
We neglect the effect of dynamical friction on the Draco dwarf as the dynamical friction time is estimated to be of the order of $\sim 1000\,{\rm Gyr}$ \citep{Cimatti+2019}.
Due to the spherical symmetry of this potential the individual components of Draco's velocity as well as its actual position on the sky are inconsequential. The \cite{zhao09} halo evolution model has the present day virial mass of the MW NFW halo, $M_{\mathrm{vir, z=0}}$, as a free parameter. Estimates of this mass in the literature vary substantially but it is generally found to be in the range $M_{\mathrm{vir, z=0}}=0.8-1.6 \times 10^{12} M_\odot$ \citep[e.g.][]{callingham19,posti19,cautun20,li20}. Hence, we integrate Draco's orbit for different choices of the present day virial mass back in time evaluating for each mass if it is consistent with ram pressure stripping being the cause for the drop in the SFR. Our criteria for this is that the drop in the SFR should (i) align with the first passage and (ii) occur while Draco is generally within the MW's virial radius. For masses where these conditions are not satisfied either ram pressure stripping would not be the main mechanism responsible for the gas loss or the stripping would have to be largely due to a local group medium rather than the MW's corona as mentioned in Section \ref{sec:sfh}.

Integrating orbits for 17 equally spaced present day virial masses in the range $0.8-1.6 \times 10^{12}$ $M_\odot$ (i.e. $5\times 10^{10}$ $M_\odot$ apart) we find three masses that satisfy these criteria: $M_{\mathrm{vir, z=0}}=9.5\times 10^{11} M_\odot$, $M_{\mathrm{vir, z=0}}=1.25\times 10^{12} M_\odot$, and $M_{\mathrm{vir,  z=0}}=1.6\times 10^{12} M_\odot$. For the each of these orbits the first passage is quite similar although at higher masses the period is shorter and the velocity higher. The time, galactocentric distance, and velocity at pericenter are shown in Table \ref{tab:orbits}. Draco has completed three, four, and five pericentric passages at the present day for the low, medium, and high MW mass potential, respectively. We choose the medium mass as our fiducial potential and hence focus on the results of the simulations based on this orbit. We discuss the two other cases, which turn out to yield largely similar results, in Section \ref{sec:mvir}.

\begin{figure}
    \centering
    \includegraphics[width=0.49\textwidth]{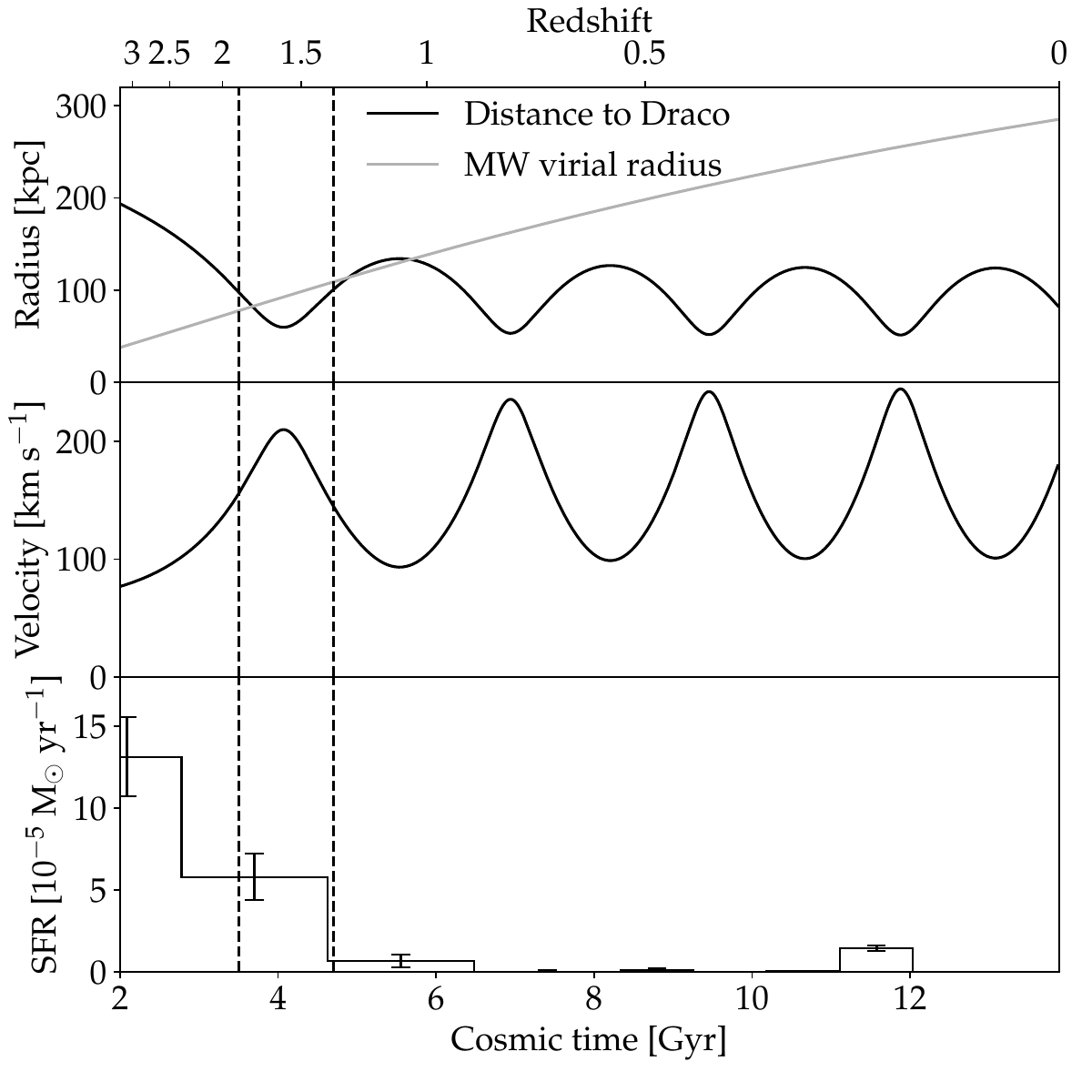}
    \caption{Top panel: Galactocentric distance to Draco along its orbit (black) and the evolution of the MW's virial radius (grey) for our fiducial potential with a present day MW virial mass of $M_{\mathrm{vir, z=0}}=1.25\times 10^{12} M_\odot$. Middle panel: Velocity of Draco along its orbit for the same potential. Bottom panel: Draco's total star formation history (see Section \ref{sec:sfh}). The dashed lines enclose the time range during the first passage that is included in the simulations.}
    \label{fig:orbit}
\end{figure}

We show the galactocentric distance and velocity of this orbit in Figure \ref{fig:orbit}, together with the SFH from Figure \ref{fig:sfh}. To assess the effect of the uncertainties in our adopted estimates for Draco's present day distance and velocity, we further integrate orbits for 1,000 different sets of these values. For these we randomly sample the distance modulus and proper motions reported in \cite{li21}, which have Gaussian uncertanties, converting them to galactocentric distances and total Galactic rest frame velocities according to Section 2.3 of that paper. We integrate these orbits with the same 17 present day virial masses between $0.8-1.6 \times 10^{12}$ $M_\odot$ as before. We find that the main effect of the uncertainties at a given virial mass is to change the orbital period. Hence, the distance and velocity evolution is mainly shifted with time while the pericentric and apocentric radii and velocities are largely similar. At our fiducial mass of $M_{\mathrm{vir, z=0}}=1.25\times 10^{12} M_\odot$, 55 per cent of the orbits have a first pericentric passage with $d_{\mathrm{GC, peri}}<R_{\mathrm{vir}}$ that occurs within a cosmic time of 3--5 Gyr. Therefore, they are consistent with the decline in SFR being due to ram pressure stripping. For the orbits that are not consistent with ram pressure stripping this is mainly due to the first pericentric passage happening too early compared to the growth of the MW's virial radius such that it occurs mostly outside of the virial radius, and so presumably outside the extent of the corona. Across our virial mass range the percentage of orbits that are consistent with ram presure stripping between cosmic times of 3--5 Gyr increases from about 25 per cent at the lowest mass to about 75 per cent at the highest mass. This increase with $M_{\mathrm{vir, z=0}}$ is due to both the virial radius growing faster and the orbital period being shorter such that more passages occur. For all of the 1,000 sets of sampled distances and velocities, between two and four of the 17 virial masses lead to passages consistent with stripping. Hence, the three orbits that we consider in the rest of this paper with the median proper motions and distances are quite representative of the space of plausible orbits given the uncertainties in the \cite{li21} data and range of reasonable MW virial masses.

\begin{table}
\centering
\noindent\begin{tabular}{l|c c c c}
\hline
  $M_{\textrm{vir, z=0}}$ & $t_{\textrm{peri}}$ & $z_{\textrm{peri}}$ & $d_{\textrm{GC, peri}}$ & $v_{\textrm{peri}}$\\
  ($10^{12}$ $M_\odot$) & (Gyr) & & (kpc) & (km s$^{-1}$) \\
   \hline
   0.95 & 4.4 & 1.7 & 62 & 200\\
   \textbf{1.25} & \textbf{4.1} & \textbf{1.6} & \textbf{59} & \textbf{211}\\
   1.6 & 3.8 & 1.5 & 56 & 222\\
\end{tabular}
\caption{Cosmic time, redshift, galactocentric distance, and total velocity in the Galactic rest frame at pericentre for the first passage for the three choices of present day MW virial masses. Our fiducial choice is highlighted in bold.}
\label{tab:orbits}
\end{table}

Given the dependence of ram pressure stripping on the square of the velocity (see eq. \ref{eq:ramstrip}), the difference between the minimum velocity at apocentre and the maximum velocity at pericentre leads to a significant, but not overwhelming, difference in the efficiency of stripping of about a factor of 4 at these points in the orbit. For many satellite galaxies this difference is much more extreme and the stripping can be assumed to occur essentially instantly at the pericentre. In our case, the difference is too small for this to be a reasonable approximation. Instead, we decide to include the parts of the orbit with velocity $v^2 > \frac{1}{2}v_{\mathrm{max}}^2$, same as was used in \cite{gatto13}. They found that including parts of the passage with less efficient stripping (i.e. with lower relative velocities) led to little difference in their inferred coronal densities. With the adopted velocity cut the simulations cover cosmic times between 3.5 and 4.7 Gyr after Big Bang. As can be seen from Figure \ref{fig:orbit}, Draco is then initially outside of the virial radius. However, it moves within the virial radius after about 200 Myr which is only about 12 per cent of the simulated time. The simulations cover galactocentric distances relatively close to the virial radius at those times in the range 60--100 kpc. Hence, the density constraints inferred from this trajectory applies to the outer parts of the early corona.

\subsection{Numerical setup}
\label{sec:nummethods}
We evolve the system using the RAMSES code \citep{teyssier02} in a three dimensional cartesian domain of $24\times 24\times 24$ kpc with Draco at the centre. We use Adaptive Mesh Refinement (AMR) with 5 levels. Each level $n$ has double the resolution of level $n-1$ ranging from $n=7$ to $n=11$ for a maximum resolution of $\Delta x\approx 11.7$ pc. Cells within a radius of $r<400$ pc from the centre of Draco's potential are always refined to the finest level. Outside of this radius cells are refined based on their mass. Cells on level $n$ are refined if they contain more than $2^{14-n} M_\odot$ corresponding to a factor 4 difference in the required density between each level.

We include optically thin radiative cooling down to $T\approx 150$K as well as heating from the metagalactic ultraviolet background (UVB). For the cooling we assume collisional ionization equilibrium using the default table in RAMSES calculated from CLOUDY \cite{ferland98}. These tables depend on density, temperature, metallicity, and ionisation fraction, which in turn depends on the redshift. For the UVB we use the model of \cite{haardt12} including the redshift evolution of its intensity during the simulations. In addition to the heating, photoionization from the UVB also affects metal line cooling rates and the mean molecular weight in the simulations. We apply the density and redshift dependent self-shielding approximation of \cite{rahmati13} to the UVB heating and photoionization calculations in each cell. That is, we multiply the photoionization and photoheating rates by a factor below unity $f_\mathrm{SSh}(n_\mathrm{H},z)$, where $n_\mathrm{H}$ is the hydrogen number density in the cell. \cite{rahmati13} found that this relation provides a tight fit to the photoionization rates in simulations with radiative transfer, with the redshift dependent parameters given in their table A1. We interpolate these parameters to the redshift at each time step during our simulations. We do however limit $f_\mathrm{SSh}$ to still allow some UV heating even in denser cells such that the equilibrium temperature is generally close to the $10^4$ K temperature of our initial isothermal profile for Draco's ISM (see Section \ref{sec:ics}). While gas is still allowed to cool below this temperature, this limitation on $f_\mathrm{SSh}$ acts as a `soft' cooling floor. Without such `soft' cooling floor, the inner parts of the initial ISM immediately starts to cool leading to a collapse to a small and very dense core. These high densities in turn lead to a short early burst of central SNe that are generally underresolved due to the high densities. We hence apply this limit to the self-shielding to avoid the strong out-of-equilibrium cooling effects, while still capturing the effect of denser gas being less affected by the metagalactic UVB. This also compensates somewhat for not including the UV heating from stars within Draco.

\subsubsection{Initial conditions}
\label{sec:ics}

\begin{table}
\centering
\noindent\begin{tabular}{l|l|c}
\hline
 Symbol & Description & Value\\
\hline
\multicolumn{3}{l}{\textbf{Milky Way}}\\
 $T_{\mathrm{cor}}$ & Coronal temperature & $2.2\times 10^6$ K \\
 $Z_{\mathrm{cor}}$ & Coronal metallicity & $0.1 Z_\odot$ \\
\multicolumn{3}{l}{\textbf{Draco}}\\
 $T_{\mathrm{ISM}}$ & ISM temperature & $10^4$ K \\
 $Z_{\mathrm{cor}}$ & ISM metallicity & 0.01 $Z_\odot$ \\
 $\rho_{0,\mathrm{DM}}$ & Central DM density & 1.12 $m_{\mathrm{p}}$ cm$^{-3}$ \\
 $r_s$ & NFW scale length & 1.46 kpc\\
 $M_{\star}$ & Stellar mass & $3.2 \times 10^5$ M$_\odot$ \\
 $r_{1/2}$ & Stellar half-mass radius & 196 pc \\
\end{tabular}
\caption{Initial parameters for the MW corona and Draco that are fixed in all simulations.}
\label{tab:fixedparams}
\end{table}

We show a summary of the initial parameters that are the same in all simulations in Table \ref{tab:fixedparams}. We initialize the coronal gas within the simulation volume to have uniform density and temperature. We choose a temperature of $T_{\mathrm{cor}}=2.2\times 10^6$ K based on observations of the MW's present day corona \citep{henley13,bregman18}. We demonstrate in Section \ref{sec:cortemp} that the value of $T_{\mathrm{cor}}$ has little effect on the stripping. In reality, an isothermal density profile in hydrostatic equilibrium with the MW potential would need to have a density that varies with radius. However, we are only concerned with the average density along the orbit, and only in the part of the orbit that we include in the simulations (see Figure \ref{fig:orbit}). In the same way, the corona need not be isothermal outside of the distances probed by this partial orbit. Thus, we are not assuming that the MW corona is isothermal everywhere to derive our lower limit on the density along part of the orbit, although we will do so in order to extrapolate this constraint to other radii in Section \ref{sec:densprof}. We do present a simulation where we vary the coronal density along the trajectory in Section \ref{sec:vardens} where we find that the changing density is indeed not important. We might expect that the early corona was less hot because the virial temperature used to be lower. The temperature of the corona is not equal to the virial temperature, which is somewhat lower at about $10^6$ K at the present day. Assuming that it is mainly comprised of shock heated accreted gas, though, it should evolve in a comparable way. For our potentials almost all of the virial temperature evolution occurs before the time at the beginning of the simulations at redshift $z=1.9$. From that time to the present day the virial temperature changes by less than 10 per cent. Hence, we expect relatively little evolution of the coronal temperature as well during this time. For the metallicity of the corona, we assume $Z_\mathrm{cor}=0.1Z_\odot$. This is somewhat less than the $\approx 0.3Z_\odot$ value preferred for the present corona \citep{miller15}, which is appropriate given the early times and relatively large distances considered (see Figure \ref{fig:orbit}).

We assume that Draco contains an isothermal ISM at temperature $T_{\mathrm{ISM}}=10^4$ K in hydrostatic equilibrium with its DM halo and stellar potential. We ignore the self-gravity of Draco's ISM because the DM dominates the mass at all radii. The stellar potential, while more important than the gravity of the gas, turned out to only very slightly reduce the stripping. But it is included since we have to assume a stellar profile for our SNIa injection (see Section \ref{sec:feedback}).We assume a spherically symmetric potential $\Phi(r)$ such that the gas density then follows
\begin{equation}
\label{eq:gasprofile}
\rho_{\mathrm{ISM}}(r) = \rho_{0,\mathrm{ISM}}\exp{\left(-\frac{\Phi(r)-\Phi_0}{c_\mathrm{s}^2}\right)},
\end{equation}
where $\rho_{0,\mathrm{ISM}}$ is the central gas density, $\Phi_0$ is the central DM+stellar potential, and $c_\mathrm{s}^2$ is the square of the isothermal sound speed. This sound speed is given by $c_\mathrm{s}^2=k_\mathrm{B} T_{\mathrm{ISM}}/(\mu m_\mathrm{p})$ where $k_\mathrm{B}$ is the Boltzman constant and $\mu$ is the mean molecular weight divided by the proton mass $m_\mathrm{p}$. Note that, due to the density dependence of UVB photoionization self-shielding, $\mu$ is generally not constant across this profile despite the constant temperature but instead tends to decrease slightly with radius. While many dwarf galaxies show signs of having a cored DM profile, Draco generally is found to be well described by a cuspy NFW profile \citep[e.g.][]{jardel13,read18,kaplinghat19,hayashi20}. Thus, we assume an NFW profile with density $\rho_{\mathrm{DM}}(r)=\rho_{0,\mathrm{DM}}(r/r_\mathrm{s})(1+r/r_\mathrm{s})^{-2}$ and potential
\begin{equation}
\Phi_{\mathrm{DM}}(r)=-\frac{4\pi G \rho_{0,\mathrm{DM}}r_\mathrm{s}^3}{r}\ln{(1+r/r_\mathrm{s})}
\end{equation}
for the DM component where we use $\rho_{0,\mathrm{DM}}=1.12$ $m_\mathrm{p}$ cm$^{-3}$ and $r_\mathrm{s}=1.46$ kpc as estimated from an extended form of spherical Jeans analysis in \cite{kaplinghat19}. For the stellar component we assume a Plummer potential
\begin{equation}
\Phi_{\star}(r)=-\frac{GM_{\star}}{\sqrt{r^2+r_{1/2}^2}},
\end{equation}
where the total stellar mass is $M_{\star}=3.2\times 10^5 M_\odot$ \citep{martin08} and the half-mass radius is $r_{1/2}=196$ pc \citep{walker07}.

In using the present day observed stellar mass and radius, we are assuming these have not evolved substantially since the beginning of our simulations at redshift $z\approx 2$. Because we begin our simulations during the relatively steep decline in the SFH this is a reasonable assumption. Indeed, according to our adopted SFH Draco formed most of its stellar mass prior to the beginning of the simulations and integrating the SFH leads to a compatible mass. In any case, as mentioned previously, the stellar potential only has a small influence on the overall potential, which is dominated by the DM.

The ISM is in pressure equilibrium with the hot gas in the corona and so extends to a radius $r_{\mathrm{ISM},0}$ where the density has dropped to $\rho_{\mathrm{ISM}}(r_{\mathrm{ISM},0})=\mu n_{\mathrm{cor}}T_{\mathrm{cor}}/T_{\mathrm{ISM}}$. Draco's potential extends further out to the tidal radius $r_t$, defined as the distance from Draco's centre towards the Galactic centre where the gravitational forces of Draco and the MW cancel out. However, Draco's potential at large radii is too weak to significantly affect the corona there. We take its change along Draco's orbit into account estimating it for each time $t$ by iteratively solving the equation of \cite{king62} (ignoring eccentricity):
\begin{equation}
r_t(t)=d_{\mathrm{GC}}(t)\left(\frac{M_{\mathrm{Dr}}(r_{\mathrm{Dr}}<r_t(t))}{3M_{\mathrm{MW}}(t,r_{\mathrm{GC}}<d_{\mathrm{GC}}(t))}\right)^{1/3},
\end{equation}
where $d_{\mathrm{GC}}(t)$ is the distance from the Galactic centre, $M_{\mathrm{Dr}}(r_{\mathrm{Dr}}<r_t(t))$ is the DM+stellar mass of Draco within the tidal radius, and $M_{\mathrm{MW}}(t,r_{\mathrm{GC}}<d_{\mathrm{GC}}(t))$ is the DM mass of the MW within Draco's distance at time $t$ (using the same evolving NFW profile as for the orbit integration). For the time span considered in our simulations $r_t$ varies within 6--13 kpc for both the low and high mass MW potential. These large tidal radii highlight that, as previously mentioned, Draco's ISM should not be affected by tidal stripping.

There are few observational estimates of the gas-phase metallicity for dwarfs around Draco's mass of $M_{\star} \approx 10^{5.5} M_\odot$. From the metallicities of the Leo P dwarf \citep{skillman13} and two dwarfs in the sample of \cite{guseva17}, as well as extrapolating the mass-metallicity relation for dwarfs of \cite{berg12}, we would expect the gas-phase metallicity for a dwarf with Draco's mass at the present day to be in the range log(O/H)$+12=7.2\pm 0.2$ corresponding to $Z_{\mathrm{ISM}}=0.02-0.05 Z_\odot$. This is an upper limit for the initial metallicity of Draco's ISM in our simulations, though, because the mass-metallicity relation evolves with redshift towards lower metallicities at fixed stellar mass \citep{huang19}. We assume $Z_{\mathrm{ISM}}=0.01Z_\odot$.

The initial ISM mass is specified by setting $\rho_{0,\mathrm{ISM}}$. We derive this by inverting a star formation law using the SFR given by Draco's SFH (see Section \ref{sec:sfh}). \cite{gatto13} used a similar approach using a steepened Kennicutt-Schmidt law with an exponent of 2.47, which has been found to better fit dwarf galaxies than the standard exponent of 1.4 \citep{roychowdhury09}. More recently, \cite{bacchini19} found that the star formation in spiral galaxies can be well fitted by a volumetric star formation law of the form
\begin{equation}
\label{eq:sflaw}
\frac{\rho_{\mathrm{SFR}}}{M_\odot \,\mathrm{yr}^{-1}\, \mathrm{kpc}^{-3}}=A\left(\frac{\rho_{\mathrm{H}}}{M_\odot \,\mathrm{pc}^{-3}}\right)^\alpha.
\end{equation}
\cite{bacchini20} showed that this law also fits well for dwarf galaxies with $A=12.59$ and $\alpha=2.03$. Such a volumetric law is more sensible for galaxies without disks compared to using surface quantities. Similar $\rho_{\mathrm{SFR}} \sim \rho_{\mathrm{ISM}}^2$ star formation laws have also been suggested in earlier work by \cite{larson69} and \cite{koeppen95}. However, the observationally derived SFR has been integrated within an elliptical region projected on the sky, rather than a sphere. Approximating the ellipsis as being circular, the integrated volume thus represents a cylinder with cylindrical radius $R<R_{\mathrm{SF}}$. Hence, we compute the SFR from the gas density within the intersection of our assumed spherical ISM distribution with radius $r_{\mathrm{0,ISM}}$ and the observed projected region with cylindrical radius $R_{\mathrm{SF}}$:
\begin{multline}
\label{eq:sfr}
\Psi_{\mathrm{tot}}(R<R_{\mathrm{SF}}) =\\ 4\pi\int_0^{R_{\mathrm{SF}}} \int_0^{\sqrt{r_{\mathrm{ISM,0}}^2+z^2}} \rho_{\mathrm{SFR}}(r=\sqrt{R^2+z^2}) R \; \mathrm{d}R \mathrm{d}z.
\end{multline}

Due to our inclusion of supernova feedback, the initial gas distribution is not in equilibrium. The feedback quickly causes the initial gas profile to expand and flatten. This lowers the SFR and, hence, the feedback. However, the expelled gas eventually falls back and reaches the centre after some tens of Myr. This causes the SFR and feedback to increase and the cycle begins anew. Due to this behaviour the SFR oscillates with an average value that is at early times a bit lower than the initial value. Therefore, the gas distribution should be initialized to a higher than observed SFR such that this early time-averaged SFR matches the observed value. We find that an initial SFR that is $f_{\mathrm{boost}}=1.2$ times the observed one leads to an average SFR over the first several cycles close to the observed value. Hence, the left hand side of equation \eqref{eq:sfr} should be multiplied by $f_{\mathrm{boost}}$. The SFR at the beginning of the simulations is then initialized to $f_{\mathrm{boost}}\Psi_{\mathrm{tot}}(R<R_{\mathrm{SF}})=f_{\mathrm{boost}}\times 5.8\times 10^{-5}$ $M_\odot$ yr$^{-1}$ in accordance with the SFH (see Figure \ref{fig:sfh}), except in Section \ref{sec:lowsfr} where we probe a lower SFR.

Using eq. \eqref{eq:sflaw} with the hydrostatic gas profile, eq. \eqref{eq:gasprofile}, and assuming a constant hydrogen mass fraction $f_{\mathrm{H}}$ then yields
\begin{multline}
f_{\mathrm{boost}}\Psi_{\mathrm{tot}}(R<R_{\mathrm{SF}}) = 4\pi A(f_{\mathrm{H}}\rho_{0,\mathrm{ISM}})^\alpha \times\\ \int_0^{R_{\mathrm{SF}}} \int_0^{\sqrt{r_{\mathrm{ISM,0}}^2+z^2}} \exp{\left[-\alpha\left(-\frac{\Phi(r)-\Phi_0}{c_s^2}\right)\right]} R \; \mathrm{d}R \mathrm{d}z
\end{multline}

and so the central density $\rho_{0,\mathrm{ISM}}$ is given by
\begin{multline}
\label{eq:rhocentral}
\rho_{0,\mathrm{ISM}} = \\
\frac{1}{f_{\mathrm{H}}}\left(\frac{f_{\mathrm{boost}}\Psi_{\mathrm{tot}}(R<R_{\mathrm{SF}})}{4\pi A \int_0^{R_{\mathrm{SF}}} \int_0^{\sqrt{r_{\mathrm{0,ISM}}^2+z^2}} \exp{\left[-\alpha\left(-\frac{\Phi(r)-\Phi_0}{c_s^2}\right)\right]} R \; \mathrm{d}R \mathrm{d}z}\right)^{1/\alpha}
\end{multline}
We assume that the hydrogen to total gas density conversion factor is $1/f_{\mathrm{H}}=1.36$. Due to the dependence of $r_{0,\mathrm{ISM}}$ on the density, and consequently on $\rho_{\mathrm{0,ISM}}$, eq. \eqref{eq:rhocentral} has to be solved iteratively.

We show the values of $\rho_{\mathrm{0,ISM}}$ and $r_{\mathrm{0,ISM}}$ for each coronal density in Table \ref{tab:varyingparams}. As can be seen, $r_{\mathrm{0,ISM}}$ is much smaller than $R_{\mathrm{SF}}=700$ pc for all coronal densities. Hence, in practice the entire gas profile is always enclosed within $R_{\mathrm{SF}}$ and eq. \eqref{eq:rhocentral} can be simplified to use an integral over a sphere of radius $r_{\mathrm{0,ISM}}$. Of course, the extent of the ISM should be at least as large as $r_{\mathrm{SF}}$ to be consistent with the SFR. However, this is still effectively the case due to the SN feedback and the effects of ram pressure which quickly cause the cold gas distribution to expand significantly. Due to this, the radius within which the density is on average much higher than the coronal density is effectively greater than $R_{\mathrm{SF}}$ in all cases shortly after the start of the simulations.

\begin{table}
\centering
\noindent\begin{tabular}{r|c c}
\hline
  $n_{0,\textrm{cor}}$ & $\rho_{0,\textrm{ISM}}$ & $r_{\textrm{0,ISM}}$\\
  (cm$^{-3}$) & ($m_p$ cm$^{-3}$) & (pc)\\
   \hline
   $5\times 10^{-4}$ & 3.96 & 270 \\
   $7\times 10^{-4}$ & 4.01 & 245 \\
   $8\times 10^{-4}$ & 4.04 & 234 \\
   $9\times 10^{-4}$ & 4.07 & 220 \\
           $10^{-3}$ & 4.10 & 210 \\
\end{tabular}
\caption{Initial central density and radius of Draco's ISM for each simulated coronal density.}
\label{tab:varyingparams}
\end{table}

\subsubsection{Velocity injection}
\label{sec:velinj}
We use a `wind tunnel' type of setup to simulate Draco moving along its orbit through the corona. That is, we inject a time-dependent velocity at the lower boundary in $y$ pointing towards the upper boundary while the other boundaries are zero-gradient (`outflow') boundaries. Section \ref{sec:orbitint} describes how we derive the velocities along the orbit $v(t)$. However, in practice this velocity needs to be amplified somewhat in order for the velocity inside the simulation domain to actually follow the wanted evolution. This is due to the fact that the velocity of the gas injected at the boundary immediately changes as it moves into the volume and either collides with or lags behind the slower or faster gas in front of it. The result is that the velocity evolution inside the volume becomes a smoothed out version of the injected velocity evolution with velocities that rise and fall too slowly in time and that are generally too low. We find that injecting the velocity according to
\begin{equation}
v_{\mathrm{inj}}(t) = v(t)\left(1+\frac{v(t)-v(t=0)}{v_{\mathrm{ref}}}\right),
\end{equation}
where $v_{\mathrm{ref}}$ depends on the potential, leads to the actual velocity inside the volume being within a few percent of $v(t)$ at all times. We find $v_{\mathrm{ref}}$ from simulations that include only the corona (i.e. does not actually contain Draco), which are very computationally inexpensive to run. For our fiducial potential $v_{\mathrm{ref}}=219$ km s$^{-1}$. For the low mass and high mass potentials discussed in Section \ref{sec:mvir} $v_{\mathrm{ref}}=206$ km s$^{-1}$ and $234$ km s$^{-1}$, respectively.

\subsubsection{Supernova feedback}
\label{sec:feedback}
Although Draco's SFR is relatively low at the epoch that we begin the simulations, feedback from supernovae can still significantly change the stripping efficiency by facilitating ram pressure stripping \citep{gatto13}. However, only the energy input is important with the added mass and metals being negligible. Because of this, we do not directly simulate the stellar population. That is, we do not include star particles representing star clusters with continuous feedback. Instead we calculate the probability of an individual SNII occurring in each cell based on its current cold gas density and use a subgrid approach to add individual SNIa according to the SFH. Due to the low SFR the injected SNe really do represent each individual SN explosion rather than the combined feedback of some population.

SNII are caused by stars with initial masses $> 8 M_{\odot}$ and for our purposes occur essentially instantly after their formation. Thus, the rate of SNII in cell $i$ is proportional to the SFR in that cell
\begin{equation}
r_{\mathrm{SNII},i}(t) = \mathrm{N}_{\mathrm{SNII}}\Psi_i(t)
\end{equation}
where N$_{\mathrm{SNII}}$ is the number of SNII per unit stellar mass formed. This depends solely on the assumed Initial Mass Function (IMF) and we adopt $0.0118$ SNII $M_\odot^{-1}$ appropriate for a \cite{chabrier03} IMF \citep{pillepich18}. We only include cells that have a temperature $T < 2\times 10^4$ K. The SFR in each cell is calculated from the density using the same volumetric star formation law of \cite{bacchini19} used in estimating the initial gas mass (eq. \eqref{eq:sflaw}, see Section \ref{sec:ics}). Given the initial SFR we expect $\sim 10$ Myr between each individual SNII occurrence anywhere in the galaxy. Because of this we estimate the number of SNII in each cell from Poisson distributions with the average number of events during the simulation time step in any cell being $\lambda \ll 1$. In practice, multiple SNe occurring in the same cell at the same time is so unlikely that it never happens during our simulations.

Unlike SNII, a star can explode as a SNIa after a substantial amount of time has passed since its birth. Consequently, the SNIa rate depends on the entire SFH rather than just the current SFR. Therefore, we cannot calculate this rate for each individual cell as we do for SNII. However, because we have the global SFH (see Section \ref{sec:sfh}) we can calculate the global rate at a given time. The global SNIa rate can be expressed as the convolution of the SFH with a \emph{delay time distribution} (DTD),
\begin{equation}
r_{\mathrm{SNIa}}(t) = \int_0^t \Psi(t-t') \dtd(t') \; \mathrm{d}t'.
\end{equation}
Unfortunately, the DTD is quite uncertain. We use the expression reported in \cite{heringer19}, $\dtd(t)=10^{-12.15} \mathrm{M}_\odot^{-1} \mathrm{yr}^{-1} (\frac{t}{1\;\mathrm{ Gyr}})^{-1.34}$ for $t>0.1$ Gyr and zero otherwise. During the simulation runs, the SFH is updated every 10 Myr with the global SFR as computed from the sum of the SFRs in all cells with cool gas that we find during the SNII rate calculation. The SNIa rates during the simulations are generally substantially lower than the SNII rates. The number of SNIa during each time step is estimated from a Poisson distribution with $\lambda$ set according to the rate as for SNII. We assume that SNIa occur within a stellar distribution that follows the observed Plummer profile also used for the stellar potential (see Section \ref{sec:ics}). Accordingly, SNIa are injected at random locations drawn from this distribution. Unlike SNII, which by definition should only occur in regions with relatively high densities, there is a small risk of SNIa being injected in very low density regions leading to extremely high velocities and numerical issues. We avoid this by not allowing SNIa that would enclose only 1 $M_\odot$ of gas to be injected. In practice, this happens extremely rarely and so the lower SNIa rate from skipping these events is not a concern.

Injecting SNe as thermal energy generally requires resolutions in excess of our standard resolution of $\Delta x = 11.7$ kpc to avoid the so-called `overcooling' problem \citep{katz92,navarro93}. Such overcooling would cause SNe to do little work on the surrounding gas and so the effect of feedback on the gas would end up being severely underestimated. To avoid this issue we instead inject the energy as a mix of thermal and kinetic energy according to the density of each cell within the initial blast. This scheme is largely similar to the `mechanical feedback' or `momentum feedback' schemes of e.g. \cite{kimm14,hopkins14,rosdahl17,hopkins18b,gentry20} who found that the evolution of a SN converged towards the correct solution at much lower resolutions than pure thermal or pure kinetic energy injection. Our implementation differs in that we do not inject any mass and we do not have any star particle associated with the SN. This simplifies the equations because the blast is always perfectly centered on a cell and occurs in the same reference frame as the gas. Additionally, we require that all cells within the initial blast be on the finest refinement level. Due to our radius and density dependent refinement criteria, an SNII occurring in, or directly next to, a cell on a coarser level would only occur extremely rarely (a couple during the entire 1.1--1.4 Gyr of the simulation). However, we disallow such SNe because they tend to cause numerical instability. Due to their extreme rarity this does not overall affect the simulations. The basis of the SN injection scheme is in each cell to either inject all the energy as kinetic energy if the mass in that cell is relatively low, or to inject a combination of thermal and kinetic energy with the kinetic energy corresponding to the `terminal momentum' otherwise. The total energy injected per SN is always $E_{\mathrm{SN}}=10^{51}$ erg in either case.

When a SN is supposed to occur in a cell, we inject it into a region centered on that cell and covering any surrounding cells sharing either a line or a plane with it. Because we assume that all the surrounding cells are on the same refinement level as the central cell, this region always covers a $3\times 3\times 3$ cube with the 8 corner cells removed, i.e. $N_{\mathrm{SN}}=19$ cells. The volume of the injection region is hence $V_{\mathrm{SN}} = 19V_{\mathrm{cell}}$ where $V_{\mathrm{cell}}$ is the cell volume. We weight the momentum and energy injected into each cell $i$. This weight is $w_i=1$ for cells directly adjacent to the central cell, i.e. sharing a plane, and $w_i=1/2$ for cells diagonally from the central cell, i.e. sharing only a line. The central cell has a weight of 1 and hence the weights sum to 13. Because the central cell has no well defined direction away from the centre for the injection momentum vector, we always inject only thermal energy in this cell. This represents only about 5 per cent of the total injection volume and so even though overcooling might occur in this cell the effect of this on the overall SN evolution is insignificant. Hence, the momentum injection covers $N_{\mathrm{mom}}=N_{\mathrm{SN}}-1=18$ cells in a volume of $V_{\mathrm{mom}}=V_{\mathrm{SN}}-V_{\mathrm{cell}}=18V_{\mathrm{cell}}$ with weights summing to 12.

The terminal momentum is the total momentum of the SN during the momentum conserving snowplow phase. This follows the radiative phase which ends once most of the thermal energy has been radiated away. Thus, the scheme alleviates overcooling in cases where the early stages of the SN cannot be resolved by initialising the SN bubble in a later evolutionary stage. If the SN \emph{can} be resolved injecting purely kinetic energy works because collisions will correctly convert this partly into thermal energy. We assume a similar expression for the terminal momentum in a given cell $i$ as \cite{rosdahl17}:
\begin{equation}
\label{eq:pterm}
p_{\mathrm{term},i} = 3 \times 10^5 \,M_\odot \;\mathrm{km} \,\mathrm{s}^{-1} \; E_{51}^{16/17} n_i^{-2/17} f_{Z,i}^{3/2},
\end{equation}
where $E_{51}\equiv E_{\mathrm{SN}}/(10^{51}$ erg), $n_i$ is the particle density in cm$^{-3}$, and
\begin{equation}
f_ {Z,i} = \begin{cases}2& Z_i < 0.01Z_{\odot}\\ (Z_i/Z_{\odot})^{-0.14}& \text{otherwise}\end{cases}.
\end{equation}
As previously mentioned, we always inject a total energy of $E_{51}=1$ with each cell receiving
\begin{equation}
\Delta E_i=\frac{w_i}{\Sigma_i w_i} E_{51}=\frac{w_i}{13} E_{51}.
\end{equation}

In the case where the energy is injected purely as kinetic energy, the total injected momentum is
\begin{equation}
\label{eq:pej}
p_{\mathrm{SN}} = f_{\mathrm{vol}}\sqrt{2M_{\mathrm{SN}} E_{51}},
\end{equation}
where $M_{\mathrm{SN}}$ is the total gas mass within the injection region and $f_{\mathrm{vol}}=V_{\mathrm{mom}}/V_{\mathrm{SN}}=18/19$ is a slight correction factor to take into account that momentum is not injected into the central cell.

An amount of momentum proportional to either eq. \eqref{eq:pej} or eq. \eqref{eq:pterm} is added to each cell in the injection region according to
\begin{equation}
\label{eq:pinj}
\Delta p_i = \frac{3w_i}{2 N_{\mathrm{mom}}} \min{(f_{\mathrm{vol}}\sqrt{2\times \frac{13}{w_i}m_i E_{51}}, \,p_{\mathrm{term},i})},
\end{equation}
where the first factor ensures that the weighting produces the correct total momentum across the 18 cells where momentum is injected, and $\frac{13}{w_i}m_i=\frac{13}{w_i}\rho_i V_{\mathrm{cell}}$ is related to the weighted mass in the cell. As the terminal momentum is weakly anti-correlated with density, at a given resolution the first argument will be smaller for low densities where the SN is resolved while the second argument will be smaller at higher densities. Conversely, for a given density, the first argument will be smaller at higher resolutions, i.e. smaller $\Delta x$, because $m_i \propto V_{\mathrm{cell}} = (\Delta x)^{3}$ while the second argument is independent of resolution and so will be the smaller argument at lower resolutions. Hence, the effect is to inject only kinetic energy when the Sedov-Taylor phase can be resolved and otherwise inject energy according to the momentum-conserving snowplow phase, as mentioned previously.

\section{Results}
\label{sec:results}
\begin{figure*}
    \centering
    \includegraphics[width=\textwidth]{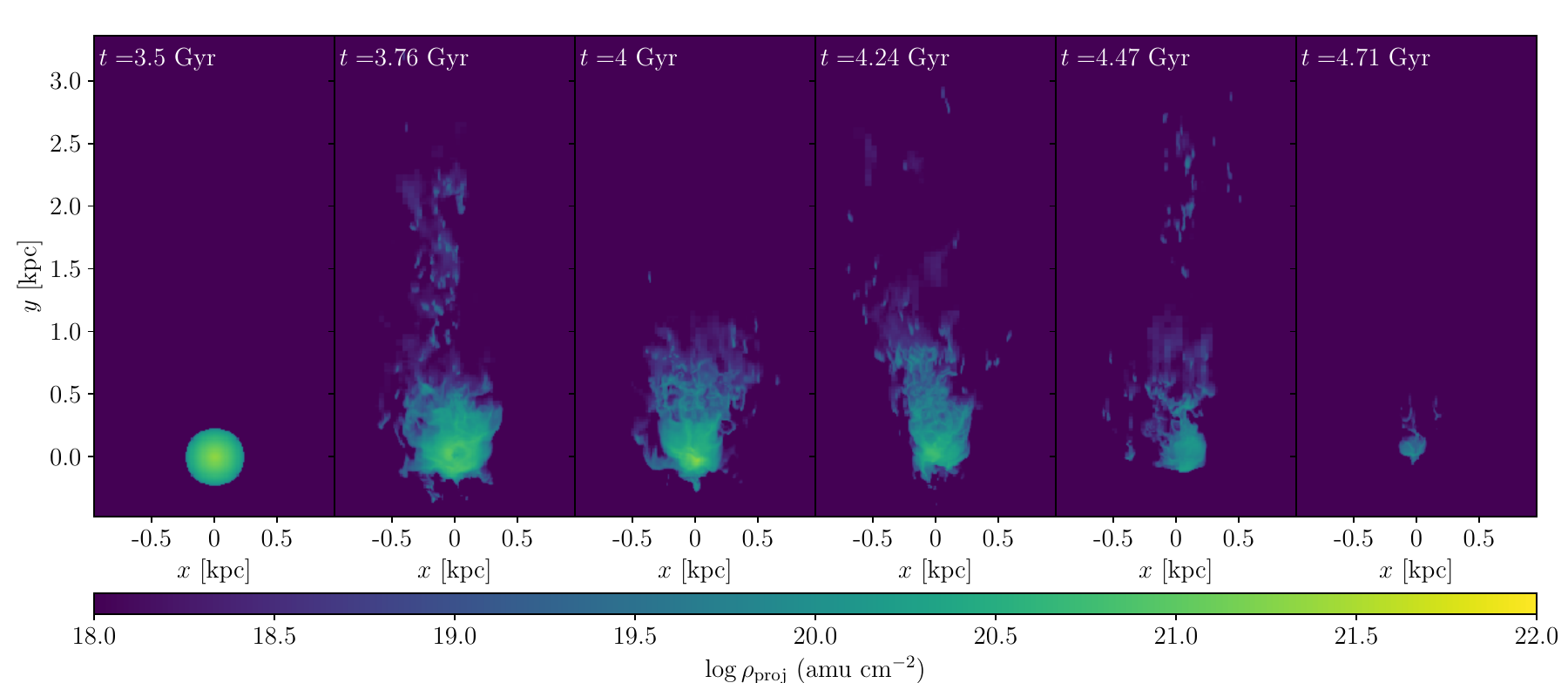}
    \caption{Projected density along the $z$ axis of gas significantly below the coronal temperature at $T<10^6$ K in the simulation with coronal density $n_{\mathrm{cor}}=8\times 10^{-4}$ cm$^{-3}$. The leftmost panel shows the initial density at the start of the simulation at a cosmic time of 3.5 Gyr. The following panels show the density at progressively later times approximately 250 Myr apart until the end of the simulation at 4.7 Gyr. Only the relevant part of the much larger $24^3$ kpc$^3$ simulation volume is shown.}
    \label{fig:colddens}
\end{figure*}

We show the evolution of the cold gas in Draco in a simulation with a coronal density of $n_{\mathrm{cor}}=8\times 10^{-4}$ cm$^{-3}$ in Figure \ref{fig:colddens}. At this coronal density essentially all of the ISM has been lost by the end of the simulation as described below. As expected, ram pressure progressively strips Draco's ISM from the outside in with the denser inner parts being lost last. There is no large tail and only small clumps of stripped gas upstream. This might be expected due to heating from the relatively intense UV background at $z>1.3$. However, in our simulations this cannot be assessed due to the poor resolution of low density stripped gas. Cells containing such gas will be on coarser refinement levels, due to our mass-based refinement criterion, and so be severely smoothed out leading to too efficient mixing with the corona. Hence, we cannot follow the evolution of this gas, although it will generally be unbound and so is not important for our analysis. The front of the ISM is pushed upstream but only by a fraction of the size of its initial distribution. Hence, the gas is lost directly to the corona rather than as a stream of cold gas that is gradually pushed out to eventually become unbound from Draco.

We show the main quantities derived from our simulations, as described in the rest of this section, in Table \ref{tab:results}.

\begin{table*}
\noindent\begin{tabular}{l|c c c c c c}
\hline
  $M_{\textrm{vir, z=0}}$ & $\langle r \rangle$ & $\langle r \rangle/R_{\mathrm{vir}}$ & $n_{\textrm{cor,min}}$ & $M_{\textrm{cor,min}}$ & $f_{\textrm{b,min}}$ & $n_{\textrm{cor,max}}$\\
  ($10^{12}$ $M_\odot$) & (kpc) & & (cm$^{-3}$) & ($M_\odot$) &  & (cm$^{-3}$) \\
   \hline
   0.95 & 76 & 0.82 & $7 \times 10^{-4}$ & $6.8 \times 10^{10}$ & 0.18 & $7 \times 10^{-4}$\\
   \textbf{1.25} & \textbf{72} & \textbf{0.76} & $\mathbf{8 \times 10^{-4}}$ & $\mathbf{7.3 \times 10^{10}}$ & \textbf{0.17} & $\mathbf{9 \times 10^{-4}}$\\
   1.6 & 69 & 0.74 & $8 \times 10^{-4}$ & $7.2 \times 10^{10}$ & 0.15 & $10^{-3}$\\
\end{tabular}
\caption{For the three possible choices of the present day MW virial mass, we show our derived constraints on the corona. The second column is Draco's average galactocentric distance $\langle r\rangle$ during first infall; this average is weighted by the mass loss (see eq. \ref{eq:ravg}). The third column is this distance as a fraction of the virial radius at the time of pericentre. The following columns are the minimum coronal density $n_{\textrm{cor,min}}$ required for complete stripping of Draco's ISM, and the corresponding minimum coronal mass and baryon fraction based on the fitted isothermal density profile (see Section \ref{sec:densprof}). The rightmost column is the upper limit on the coronal density $n_{\textrm{cor,max}}$ derived from the maximum baryon fraction $f_{\textrm{b,max}}=f_\mathrm{b}=0.18$. The row of our fiducial virial mass is highlighted in bold.}
\label{tab:results}
\end{table*}

\subsection{Density of the early MW corona}
\subsubsection{Average coronal density along Draco's first infall at $z=1.3-1.9$}
\label{sec:cordenshiz}
We use the stripping of Draco's ISM to put a lower bound on the average density of the corona at the times (cosmic ages of 3.5--4.7 Gyr or redshifts around $z\approx 1.3-1.9$) and distances (59--103 kpc or 0.6--1.3$R_{\mathrm{vir}}$) of the simulations. Our constraint is a lower bound because Draco's SFH does not have the time resolution needed to assess when during the passage the gas was lost completely. The mean distance is skewed towards lower distances due to the flattening in the distance evolution around the pericentre and is around 76 kpc. Despite the distance range extending beyond the virial radius Draco is generally within the virial radius due to its growth during the simulations and the mean distance in terms of the virial radius is $0.81R_{\mathrm{vir}}$.

\begin{figure}
    \centering
    \includegraphics[width=0.49\textwidth]{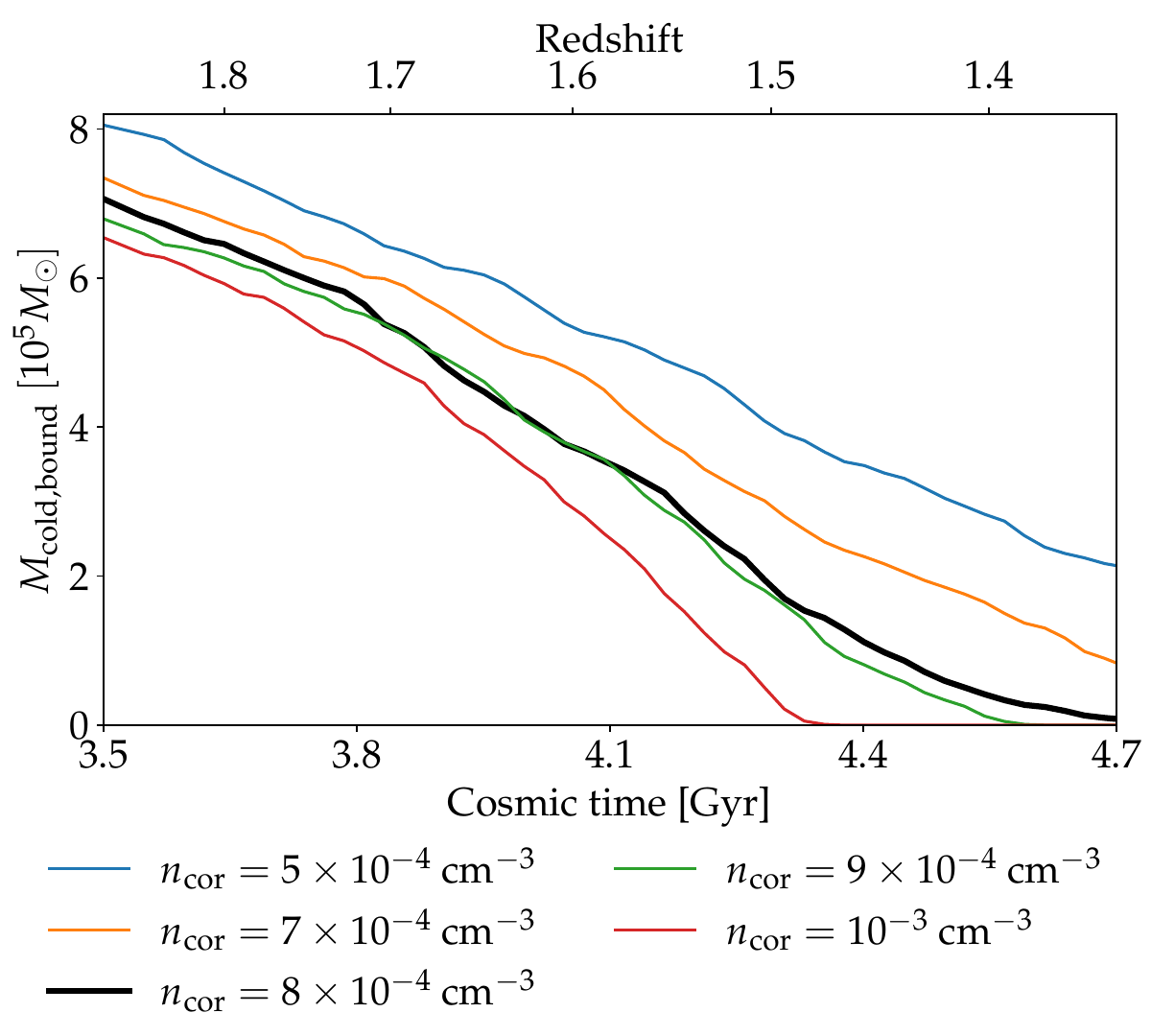}
    \caption{Evolution of the mass of cold bound gas for our simulations with the fiducial potential. The time of pericentre is in the middle at 4.1 Gyr. Complete stripping occurs for $n_{\mathrm{cor}}\geq 8\times 10^{-4}$ cm$^{-3}$.}
    \label{fig:mcoldbound}
\end{figure}

We quantify the remaining mass of Draco's ISM during the simulation as the total mass of cold gas that is gravitationally bound to Draco $M_{\mathrm{cold, bound}}$. We define `cold' as $T<10^5$ K and bound as having a total velocity less than the escape velocity to infinity $|\mathbf{v}|<v_{\mathrm{esc}}(r)=\sqrt{2|\Phi(r)|}$. The criterion that the gas be bound does not exclude a significant mass of gas not already excluded by the temperature criterion, due to most stripped gas being quickly heated as previously mentioned. The initial value of $M_{\mathrm{cold, bound}}$ differs by about 25 per cent from the lowest to the highest coronal density due to the initial pressure equilibrium leading to smaller initial ISM radii at higher coronal densities (see Table \ref{tab:varyingparams}). However, this has essentially no effect on the early SFR which is initially dominated by the denser gas near the centre. We show the evolution of $M_{\mathrm{cold, bound}}$ for simulations with different $n_{\mathrm{cor}}$ in Figure \ref{fig:mcoldbound}.

As can be seen, the mass loss generally increases monotonically with coronal density as expected. The simulation with $n_{\mathrm{cor}}=8\times 10^{-4}$ cm$^{-3}$ is close to the simulation with $n_{\mathrm{cor}}=9\times 10^{-4}$ cm$^{-3}$ but complete stripping does occur slightly before the end of the simulation in the latter case. However, the otherwise relatively high sensitivity of the stripping on the density means that the ISM is clearly lost before the end of the simulation for densities $n_{\mathrm{cor}}>10^{-3}$ cm$^{-3}$ and too late for $n_{\mathrm{cor}}<7\times 10^{-4}$ cm$^{-3}$. We are hence able to, within the assumptions of the simulations, derive an accurate lower bound of $n_{\mathrm{cor,min}}$ on the coronal density at the above mentioned distances and ages of $\langle n_{\mathrm{cor,min}}(59-103$ kpc$)\rangle = 8 \times 10^{-4}$ cm$^{-3}$. At this coronal density essentially all of the ISM has been lost by the end of the simulation, with only $M_{\mathrm{cold, bound}}\approx 7500 M_\odot$, or about 1 per cent of its initial mass, remaining. As expected, the mass loss is generally greater during the middle part of the simulations where Draco's velocity is higher compared to at the beginning and end of the simulations. However, instantaneous stripping at the pericentre, as is often assumed, is clearly not a very accurate approximation. Due to this evolution in the mass loss the average distance for the purposes of our lower density limit is skewed slightly further towards lower distances than the unweighted mean distance of 76 kpc. Weighting the distance by the mass loss we find
\begin{equation}
\label{eq:ravg}
\langle r \rangle = \frac{\int_{3.5\,\mathrm{Gyr}}^{4.7\,\mathrm{Gyr}} r \frac{dM}{dt} \, dt}{\int_{3.5\,\mathrm{Gyr}}^{4.7\,\mathrm{Gyr}} \frac{dM}{dt} \, dt}\approx\frac{\int_0^{M_0} r \, dM}{M_0}=72\,\mathrm{kpc},
\end{equation}
where $M$ here is shorthand for $M_{\mathrm{cold,bound}}$ and $M_0$ is the initial value at a cosmic time of 3.5 Gyr ($z=1.9$). Thus, the $59-103$ kpc range for our lower density bound $n_{\mathrm{cor,min}}$ is centred at 72 kpc and we have $n_{\mathrm{cor,min}}(72$ kpc$) = 8 \times 10^{-4}$ cm$^{-3}$. This corresponds to $0.76R_{\mathrm{vir}}$.

\cite{gatto13} found that the analytical estimate of the lower limit on the coronal density from eq. \eqref{eq:ramstrip} was significantly less than the result of their simulations for Carina and Sextans. If we plug $v_{\mathrm{peri}}=211$ km s$^{-1}$, the initial average density of Draco's ISM in our simulations of $n_{\mathrm{ISM}}=0.48$ cm$^{-3}$, and Draco's measured present day stellar velocity dispersion of $\sigma_\star=9.1$ km s$^{-1}$ \citep{simon19} into eq.~(\ref{eq:ramstrip}) we obtain $n_{\mathrm{cor}} \gtrsim 9 \times 10^{-4}$ cm$^{-3}$ in remarkably good agreement with our simulation results. We caution though that, as mentioned previously, our simulations disagree with the assumption of instantaneous stripping at pericentre used in the crude analytical estimate and so the close agreement must be partially a coincidence. It is unlikely that eq. \eqref{eq:ramstrip} agrees this closely with simulations in general for the range of different masses and SN rates of MW satellites.

\subsubsection{Density profile and mass of the corona at $z\approx 1.6$}
\label{sec:densprof}
Based on our lower bound on the coronal density we can derive lower bounds for the density profile, $n_{\mathrm{cor,min}}(r)$, and the total mass, $M_{\mathrm{cor,min}}$, of the MW corona. In order to do this we have to assume a profile that can be fit from our single point of $n_{\mathrm{cor,min}}(72$ kpc$) = 8 \times 10^{-4}$ cm$^{-3}$. We assume the same hydrostatic isothermal density profile as for Draco's ISM, eq. \eqref{eq:gasprofile}, with the same NFW potential as used for the orbit integration and temperature $T_{\mathrm{cor}}=2.2\times 10^6$ K as used in the simulations. At this temperature self-shielding is unimportant and so $\mu\approx 0.6$ regardless of the density. This profile is expected to be too flat very close to the centre at $r\lesssim 0.1R_{\mathrm{vir}}$ because it ignores the disc and bulge potential, which are uncertain at high redshift. However, this is not important because our constraint applies to much larger distances and the effect on the overall estimated mass of the corona from this small fraction of its volume is not significant. In any case, while a hydrostatic profile is often used for the corona in the literature, it can only be a rough approximation given the presence of in and -outflowing gas in the corona. We discuss this in Section \ref{sec:gasdist}. Due to the evolution of the MW potential, we have to choose a specific time to fit the profile. We choose the time at pericentre, i.e. a cosmic age of $t=4.1$ Gyr corresponding to a redshift of $z=1.6$, which is also the time halfway through the simulations and within a few per cent of the mass loss weighted time, as a natural choice. At this time, this MW model has a scale density of $\rho_{0,\mathrm{DM}}=0.2$ $m_\mathrm{p}$ cm$^{-3}$ and a scale radius of $r_s=19.9$ kpc. We find that the central density of the MW corona has to be at least $n_{0,\mathrm{cor}}\approx 5.7\times 10^{-3}$ cm$^{-3}$ for the density at the mass loss weighted mean distance of Draco's orbit of 72 kpc to be above the $8\times 10^{-4}$ cm$^{-3}$ lower density bound.

\begin{figure}
    \centering
    \includegraphics[width=0.49\textwidth]{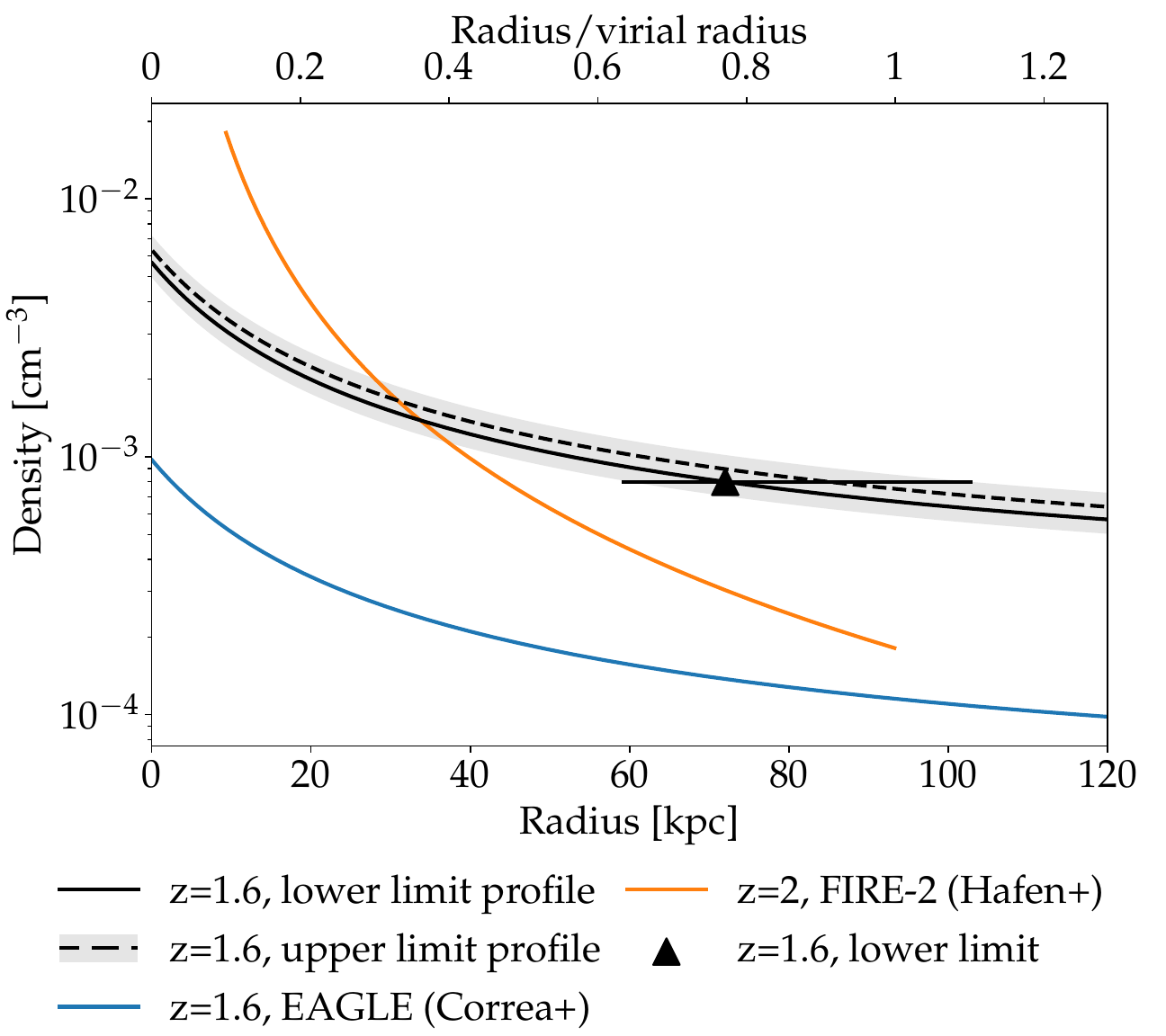}
    \caption{Our estimated lower bound on the density of the corona at redshift $z\approx 1.6$ (triangle) with the range of distances of Draco's orbit indicated by the horizontal bars. The black lines are isothermal density profiles representing the lower and upper bounds for the corona. The lower bound profile (solid line) is fitted to our lower bound estimate. The upper bound (dashed line) is based on the cosmological baryon fraction (see Section \ref{sec:upperbound}). The grey band surrounding the dashed line shows the wider range of upper limits considering the uncertain ISM mass. The orange line is the $n\propto r^{-2}$ profile fit to the average density of the gas within the haloes of MW like progenitors at $z=2$ in the FIRE-2 simulations of \protect\cite{hafen19}. The blue line is the isothermal density profile based on the mass from eq. \eqref{eq:mcorcorrea} fitted to the EAGLE simulations in \protect\cite{correa18}.}
    \label{fig:densprofz16}
\end{figure}

We show our estimate on the lower bound together with this isothermal profile in Figure \ref{fig:densprofz16}. We also show our estimated upper bound isothermal density profile derived from the cosmological baryon fraction as described in Section \ref{sec:upperbound}. As can be seen, these profiles are very close. To our knowledge, there are no observations or previous observationally based estimates of the hot coronal gas in MW progenitor mass galaxies at these redshifts to compare with. However, there are some theoretical predictions from galaxy simulations. \cite{hafen19} examined the evolution of the gas at $0.1 < r/R_{\mathrm{vir}} < 1$ for the progenitors of (among others) MW mass galaxies in the FIRE-2 cosmological zoom-in simulations \citep{hopkins18}. They found that the density at $z=2$ followed the $n\propto r^{-2}$ profile shown in orange in Figure \ref{fig:densprofz16}. As this is at a slightly higher redshift it is an upper limit on the density at $z=1.6$ due to the fact that the density of the corona decreases with time. Additionally, this profile is for gas at all temperatures within the halo and only roughly half of the mass of this gas is in the hot ($T>10^{5.3}$ K) phase. In any case, the profile is steeper than the isothermal NFW profile with densities above the upper isothermal limit near the centre and about a quarter of our lower bound at $r=0.76R_{\mathrm{vir}}$. In addition to the FIRE-2 result, we also show the isothermal profile derived from the coronal mass given by \cite{correa18}. They examined hot coronae within haloes at $0.15R_{200}<r<R_{200}$ in the EAGLE cosmological simulations \citep{schaye15}, where $R_{200}$ is the radius within which the mean DM density is 200 times the critical density and is slightly less than the virial radius. They found that this hot gas mass normalised by the total baryonic mass, estimated as $f_\mathrm{b} M_{200}$, for $z<2$ is approximately given by
\begin{multline}
\label{eq:mcorcorrea}
\log{\left(\frac{M_{\mathrm{cor}}}{f_\mathrm{b} M_{200}}\right)}=-0.79 + 0.31\tilde{z} - 0.96\tilde{z}^2 +\\
(0.52 - 0.57\tilde{z} + 0.85\tilde{z}^2)x - 0.05x^2,
\end{multline}
where $\tilde{z}\equiv\log{(1+z)}$, $x\equiv\log{(M_{200}/10^{12} M_\odot)}$, and $M_{200}$ is the DM mass within $R_{200}$. As can be seen, this leads to a low mass corona that only accounts for a small fraction of the baryonic mass. However, this depends strongly on feedback which can both help the corona grow by expelling gas from the ISM and hinder its growth by removing gas from the corona (in addition to the effect of even lowering the infalling baryon fraction in some cases as mentioned in Appendix \ref{sec:coronaevol}). They find that stellar feedback mainly causes the former while AGN feedback mainly leads to the latter effect. By comparing a few different EAGLE simulations with different subgrid feedback implementations they find that the corona of a MW mass halo at $z=0$ can contain anywhere from 4 to 40 per cent of $f_\mathrm{b} M_{200}$. This is substantially below our lower limit that is consistent with containing all the `missing' baryons (see Section \ref{sec:upperbound}). This suggests that the strong ejective feedback in these simulations is overly effective at removing coronal gas. 
We note that this conclusion could change if our assumptions that the corona is smooth and roughly in hydrostatic equilibrium at $z\approx1.6$ were removed. 
We can, in fact, envision that in a more structured medium \citep[e.g.][]{Hummels+2019} the stripping could occur in denser regions and the average density be lower than estimated here.
However, observations of giant Ly$\alpha$ nebulae around redshift $z\sim 3$ quasars in $10^{12} M_\odot$ halos do indicate that galaxies in that regime also have massive hot coronae with similar properties to that estimated here and significantly denser than predicted by cosmological simulations \citep{pezzulli19}.

Integrating our lower limit density profile out to the virial radius at $t=4.1$ Gyr ($z=1.6$) of $R_{\mathrm{vir}}=94$ kpc we obtain an estimate of the lower bound on the mass of the corona at that time of $M_{\mathrm{cor,min}} = \int_0^{R_{\mathrm{vir}}} 4\pi r^2\mu m_\mathrm{p} n_{\mathrm{cor,min}}(r)  \, dr = 4.3\times 10^{10} M_\odot$. The lower limit on the overall average density of the entire corona $\langle n_{\mathrm{cor,min}}\rangle=3M_{\mathrm{cor,min}}/(4\pi\mu m_\mathrm{p} R_{\mathrm{vir}}^3)$ turns out to be equal to our estimate at 72 kpc i.e. $\langle n_{\mathrm{cor,min}}\rangle \approx 8\times 10^{-4}$ cm$^{-3}$.

\subsubsection{Baryon fraction and upper bounds}
\label{sec:upperbound}
To estimate the lower bound of the baryon fraction at $z=1.6$ we need to find $M_\mathrm{b}\approx M_{\star}+M_{\mathrm{ISM}}+M_{\mathrm{cor}}$ at this time. From the stellar mass evolution model of \cite{vandokkum13} we find that the stellar mass of the MW around the time of pericentre was $M_{\star}(z=1.6)\approx 2\times 10^{10} M_\odot$ or 40 per cent of the present day stellar mass. The mass of the MW's ISM at $z=1.6$ is highly uncertain but generally gas mass fractions were higher in the past \citep{santini14,scoville17}. Assuming that the mass fraction of the ISM $f_{\mathrm{ISM}}=M_{\mathrm{ISM}}/(M_{\star}+M_{\mathrm{ISM}})$ was about double its present value of $f_{\mathrm{ISM}}\approx 0.15$ \citep{bland-hawthorn16,mcmillan17}, as suggested by the model of \cite{dave12}, we obtain $M_{\star}(z=1.6)+M_{\mathrm{ISM}}(z=1.6)\approx 3\times 10^{10} M_\odot$. The virial mass of the MW's DM profile at $z=1.6$ in our assumed model is $4.3\times 10^{11} M_\odot$. Our determination of $M_\mathrm{cor}$ leads then to a lower bound on the baryon fraction of
\begin{equation}
    f_{\mathrm{b,min}}(z=1.6)=\frac{M_\mathrm{b}(z=1.6)}{M_{\mathrm{vir}}(z=1.6)} = \frac{7.3\times 10^{10} M_\odot}{4.3\times 10^{11} M_\odot} = 0.17.
\end{equation}
This is very close to the cosmological baryon fraction of $f_\mathrm{b}=0.18$ but still consistent with the requirement that the MW should not have a higher baryon fraction than this. Without the coronal gas the baryon fraction would only be 0.07. Hence, most of the baryons and essentially all of the `missing baryons' at $z=1.6$ would be in the corona.

We derive an upper limit on the mass of the corona based on the requirement that the baryon fraction of the MW should not exceed the cosmological baryon fraction. This limit is $M_{\mathrm{cor}}<4.8\times 10^{10} M_\odot$, about 10 per cent above the lower limit on $M_{\mathrm{cor}}$ from integrating the isothermal profile in the previous section.

From this upper limit on the coronal mass we can derive an upper limit isothermal density profile. We find that this mass is exceeded for profiles with central density $n_{0,\mathrm{cor}}> 6.4\times 10^{-3}$ cm$^{-3}$ corresponding to a density at 72 kpc of about $9\times 10^{-4}$ cm$^{-3}$. Due to the considerable uncertainty on the ISM mass of the MW at $z=1.6$ we also show a wider band in grey surrounding the upper limit in Figure \ref{fig:densprofz16}. This band is derived from upper limits considering a wide range of possible $f_{\mathrm{ISM}}$ from 0.15 to 0.5. The range of densities between our lower and upper limits is thus very narrow, and overlaps when considering the grey band. The lower limit at 72 kpc from our simulations is considerably more robust than the upper limits, though. Unlike the upper bounds, it does not assume that the entire corona, including the inner parts, follows the same isothermal density profile and does not involve any assumptions about the stellar and ISM mass of the MW.

In summary, our results suggest that the MW at $z=1.6$ had a baryon fraction close to the cosmological baryon fraction.

\subsection{The present day corona}
\label{sec:presentcorona}
\subsubsection{Literature constraints}
Unlike at high redshift, there are a number of constraints on the density of the present day MW corona in the literature. Hence, we can assess the evolution of the corona by comparing our constraint with these. We will consider the density constraints of \cite{anderson10}, \cite{gatto13}, \cite{salem15}, and \cite{putman21}. Our lower limit of $8\times 10^{-4}$ cm$^{-3}$ at $z=1.6$ is above almost all of these estimates, even though those are generally in the inner part of the halo. This highlights that the density of the corona has generally decreased significantly. This is expected as the volume within $R_{\mathrm{vir}}$ has increased by a factor of $\approx 30$ while the virial mass has increased by a factor of $\approx 3$. Hence, any volume filling medium that extends to $R_{\mathrm{vir}}$ must have become less dense on average, despite having increased in mass, for reasonable growth rates.

The constraint of \cite{anderson10} is based on the dispersion measure of pulsars in the Large Magellanic Cloud (LMC). This is an upper limit because they do not include any contribution to the dispersion measure from warm gas in the LMC. The other constraints are based on ram pressure stripping with \cite{gatto13} and \cite{salem15} being derived from simulations and \cite{putman21} being more crude lower limits calculated from eq. \eqref{eq:ramstrip}. We bin the data of \cite{putman21}, who estimated lower limits for 36 galaxies within the virial radius, by distance so as to not clutter the figure in the following subsection. However, we exclude their highest found limit for Fornax, which is an outlier at $n\gtrsim 10^{-3}$ cm$^{-3}$. As the authors point out, the assumption of instantaneous stripping at the pericentre in eq. \eqref{eq:ramstrip} probably leads to a severe overestimate for Fornax due to its high mass and relatively circular orbit. We also exclude Tucana III and Sagittarius, which have clearly undergone severe tidal stripping. The remaining 33 galaxies that we consider cover distances from 17 to 182 kpc. We use three equally spaced bins between 0 and 200 kpc, containing 19, 7, and 7 galaxies each, in order of increasing distance. The most constraining (i.e. highest) lower limits in each bin are Sculptor, Carina, and Leo II, in order of increasing distance. \cite{putman21} also estimated the minimum density needed to strip Draco finding a lower estimate than ours of $n_{\mathrm{cor,min}}=3.1^{+1.2}_{-0.9} \times 10^{-4}$ cm$^{-3}$. This is expected because they assume that the stripping occurred during the last pericentric passage at which the velocity was substantially higher than during the first (see Figure \ref{fig:orbit}). Their assumed mean density for Draco's ISM of 0.37 cm$^{-3}$ is not so different from the initial mean density in our simulations of 0.48 cm$^{-3}$, however. As shown in Section \ref{sec:cordenshiz}, plugging in the pericentric velocity and mean ISM density in our simulations the estimate from eq. \eqref{eq:ramstrip} is remarkably close to the result of our simulations. Other galaxies in \cite{putman21} that ended their star formation early, possibly around the same time as Draco, are Ursa Minor and Sculptor \citep{carrera02,dolphin03,deboer12b,savino18,betinelli19}. If they were also stripped in the earlier denser corona, this could explain why the minimum coronal densities for these derived by \cite{putman21}, especially that of Sculptor at $4.8_{-1.3}^{+1.4}\times 10^{-4}$ cm$^{-3}$, are relatively high. However, like Draco, they should also not be directly compared to our $z=1.6$ density profile because the given distances and the pericentric velocities used in eq. \eqref{eq:ramstrip} are based on their most recent pericentric passage which would then not be correct. In any case, there are many other galaxies in \cite{putman21} with lower limit coronal densities of a few $\times 10^{-4}$ cm$^{-3}$, so excluding these galaxies would not decrease their overall limits by much. We do not include the lower limits of \cite{grcevich09} because these have essentially been superseded by \cite{putman21} who use the same method but with newer, more accurate observational data (in particular the proper motions from \emph{Gaia}). In any case, the overall lower limit of \cite{grcevich09} of a few $\times 10^{-4}$ cm$^{-3}$ at $r \lesssim 150$ kpc agrees well with \cite{putman21}. The constraint of \cite{salem15} is generally lower than the other estimates at that distance of \cite{gatto13} and \cite{putman21}. The density along the orbit of the LMC could be lower than in the other parts of the corona as spherical symmetry is in any case only a rough approximation. \cite{tepper-garcia15} does find that a higher value of $n \gtrsim 2\times 10^{-4}$ cm$^{-3}$ is needed in order to explain the H$\alpha$ emission from the Magellanic Stream (MS), in agreement with the other constraints. However, the  recent simulations of \cite{lucchini21} suggests that the MS is much closer to the disc than the LMC with an average distance of $\approx 25$ kpc. \cite{stanimirovic02} estimated upper limits on the density of the corona around clouds in the MS based on pressure equilibrium. They found that the coronal density had to be less than $10^{-3}$ cm$^{-3}$ for clouds at $r=15$ kpc and less than $3\times 10^{-4}$ cm$^{-3}$ for clouds at 55 kpc. If the distance of the MS is as suggested by \cite{lucchini21} then only the former limit, which is similar to that of \cite{anderson10}, applies. Hence, we do not consider this constraint in our comparison.

\begin{figure}
    \centering
    \includegraphics[width=0.49\textwidth]{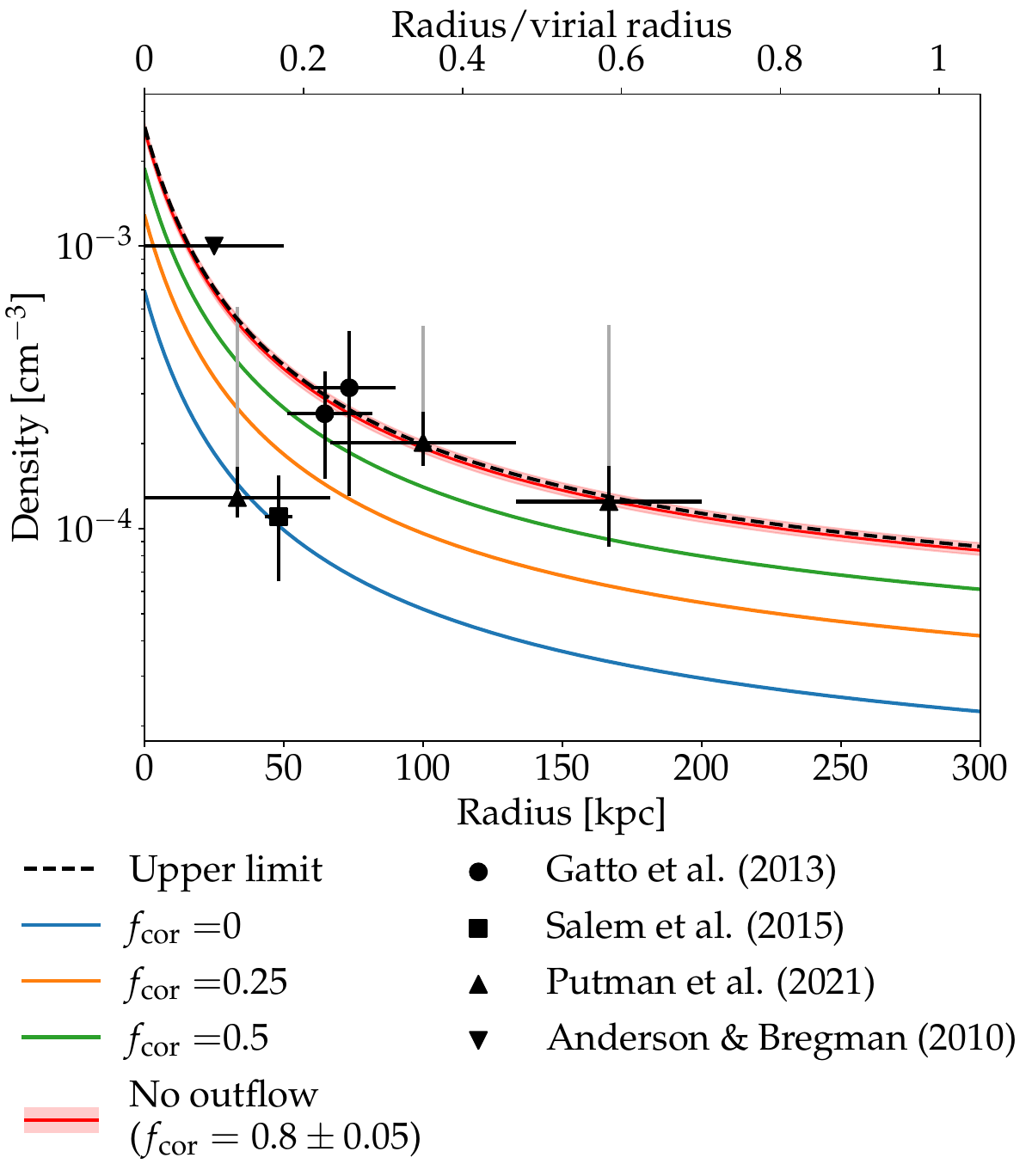}
    \caption{Isothermal density profiles representing the lower and upper bounds for the present day corona. The lower bounds have been evolved from our $z\approx 1.6$ estimate based on different assumed constant values of the fraction of gas accreted onto the halo that ends up in the corona $f_\mathrm{cor}$. The upper bound (dashed line) is based on the cosmological baryon fraction. This happens to be similar to the upper limit on the evolution of the lower limit from assuming no outflow from the corona to the IGM given by $f_\mathrm{cor}=0.8\pm 0.05$. The profiles with $f_\mathrm{cor}\geq 0.25$ agree reasonably well with most of the estimates from the literature shown: \protect\cite{gatto13} (lower and upper limits, circles), \protect\cite{salem15} (square), \protect\cite{putman21} (binned lower limits with highest lower limit in grey [see text], upwards filled triangles), and \protect\cite{anderson10} (upper limit, downwards triangle)}
    \label{fig:densprofz0}
\end{figure}

\subsubsection{Density profiles extrapolated to $z=0$}
\label{sec:densprofz0}
We show the isothermal profiles based on extrapolating our $z=1.6$ profile to the present day in Figure \ref{fig:densprofz0}, together with the literature constraints mentioned above. For the binned \cite{putman21} constraints, we show the mean and its uncertainties for each bin in black and the most constraining (i.e. highest) lower limit in each bin in grey. We have extrapolated the mass in the corona according to eq. \eqref{eq:deltamcorfb} assuming different constant values of the $f_\mathrm{cor}$ parameter. This parameter is the net mass growth of the corona, i.e. inflows from the IGM and ISM minus outflows to the IGM and ISM, normalized by the mass inflow from the IGM considered over long time scales (see Appendix \ref{sec:coronaevol}). As described in the Appendix, we exclude the scenarios where the corona is being depleted ($f_{\mathrm{cor}}<0$) or the disc is being depleted ($f_{\mathrm{cor}}>1$) based on the sustained star formation of the MW. Hence, $f_{\mathrm{cor}}$ represents the fraction of infalling gas that is not later removed from the halo and ends up in the corona, in this case from $z=1.6$ to $z=0$. The mass of the present day corona for a given value of $f_\mathrm{cor}$ is then given by
\begin{multline}
\label{eq:mcorgrowth}
M_{\mathrm{cor}}(z=0)\approx M_{\mathrm{cor}}(z=1.6) + f_\mathrm{cor} f_\mathrm{b}\Delta M_{\mathrm{vir}}=\\
4.3\times 10^{10} M_\odot + 1.5\times 10^{11} M_\odot f_\mathrm{cor}.
\end{multline}
Note that, although we add to the initial coronal mass at $z=1.6$ under the assumption that it has grown (i.e. $f_{\mathrm{cor}}>0$) to find the present day mass, this does not imply that all the gas that was present in the corona at $z=1.6$ is still in the corona at $z=0$. It is possible that a large fraction of this gas has been either accreted onto the disc or ejected from the corona since $z=1.6$ and been replaced with infall from the IGM.
For these profiles, we use our assumed present day virial mass of $1.25\times 10^{12} M_\odot$ which has a virial radius of 285 kpc and has grown by a factor of 2.85 since $z=1.6$ in the \cite{zhao09} model. We also show the upper limit profile from the cosmological baryon fraction derived from the estimated sum of the present day stellar and ISM masses of $6\times 10^{10} M_\odot$ \citep{bland-hawthorn16}. This profile has a central density of $n_{0,\mathrm{cor}}=2.7\times 10^{-3}$ cm$^{-3}$. While this is about half of the central density of our lower limit density profile at $z=1.6$, the overall average density within the virial radius in all the allowed $z=0$ coronae are even lower compared to at $z=1.6$. This is because the profiles at $z=0$ are much steeper due to the evolution in the concentration of the halo. This might seem to imply the unintuitive evolution that coronal gas has moved outwards while the potential has steepened. However, as mentioned previously, gas flows between the corona and the ISM and IGM means that the coronal gas can be largely replaced over long time scales, rather than the original more compact corona having to continually expand with the virial radius. This strong evolution in the density of the corona has important implications for studies of ram pressure stripping within it. For instance, \cite{emerick16} concluded that low mass MW satellites around $M_\star\sim 10^5 M_\odot$, just a bit smaller than Draco, were difficult to effectively ram pressure strip based on a density of $10^{-4}$ cm$^{-3}$, representative of the present day corona. As pointed out by \cite{putman21}, most of these smaller dwarfs being stripped by the denser early corona could alleviate this issue.

An upper limit on the growth of the corona can be derived by assuming no outflow from the corona to the IGM. The growth of the corona is then simply the baryons accreted onto the halo minus the growth of the disc and so the corresponding $f_\mathrm{cor,max}$ is given by
\begin{equation}
f_\mathrm{cor,max} = \frac{f_\mathrm{b}\Delta M_{\mathrm{vir}} - \Delta M_\star - \Delta M_\mathrm{ISM}}{f_\mathrm{b}\Delta M_{\mathrm{vir}}}.
\end{equation}
The growth of the stellar mass is relatively well constrained to be $\Delta M_{\star}\approx 3\times 10^{10} M_\odot$ \citep{vandokkum13} while the gas mass is a lot more uncertain but also generally lower. Using the same range of values for the ISM mass at $z=1.6$ as in Section \ref{sec:upperbound} and assuming a present day mass of $10^{10} M_\odot$ yields $f_\mathrm{cor,max}=0.8\pm 0.05$. This is consistent with the estimates of the \emph{instantaneous} fraction of accreted gas that reaches the disc $\zeta \approx 0.3-0.5$ from analytical models and cosmological simulations (see Appendix \ref{sec:coronaevol}) given that $f_\mathrm{disc}\leq \zeta$. The corresponding density profile, shown in red in Figure \ref{fig:densprofz0}, overlaps with the upper limit range from the cosmological baryon fraction. This is not surprising because the lower limit at $z=1.6$ was already close to the upper limit at that time and $f_\mathrm{cor,max}$ is the most extreme possible extrapolation. The literature constraints do prefer a relatively massive corona although due to the considerable uncertainties all values $0.5 \lesssim f_\mathrm{cor} \leq f_\mathrm{cor,max}$ are in decent agreement. This allows for outflows from the corona with $f_\mathrm{out}\lesssim 0.3$. If the baryon fraction of the matter accreted onto the halo is below $f_\mathrm{b}$ this upper limit on $f_\mathrm{out}$ becomes lower. Our range of plausible $f_\mathrm{cor}$ is well above the $f_\mathrm{cor}=0.2$ found in EAGLE \citep{correa18} which we also found to predict a too low coronal mass at $z=1.6$ compared to the lower limit mass from our density constraint.

In any case, the lower limit profiles extrapolated from our $z=1.6$ constraint are in good agreement with the present day constraints and at or below the upper limit from the cosmological baryon fraction for reasonable values of $f_\mathrm{cor}$.

\section{Discussion}
\label{sec:discussion}
\subsection{Effect of feedback on stripping}
\label{sec:effectoffb}
\begin{figure}
    \centering
    \includegraphics[width=0.49\textwidth]{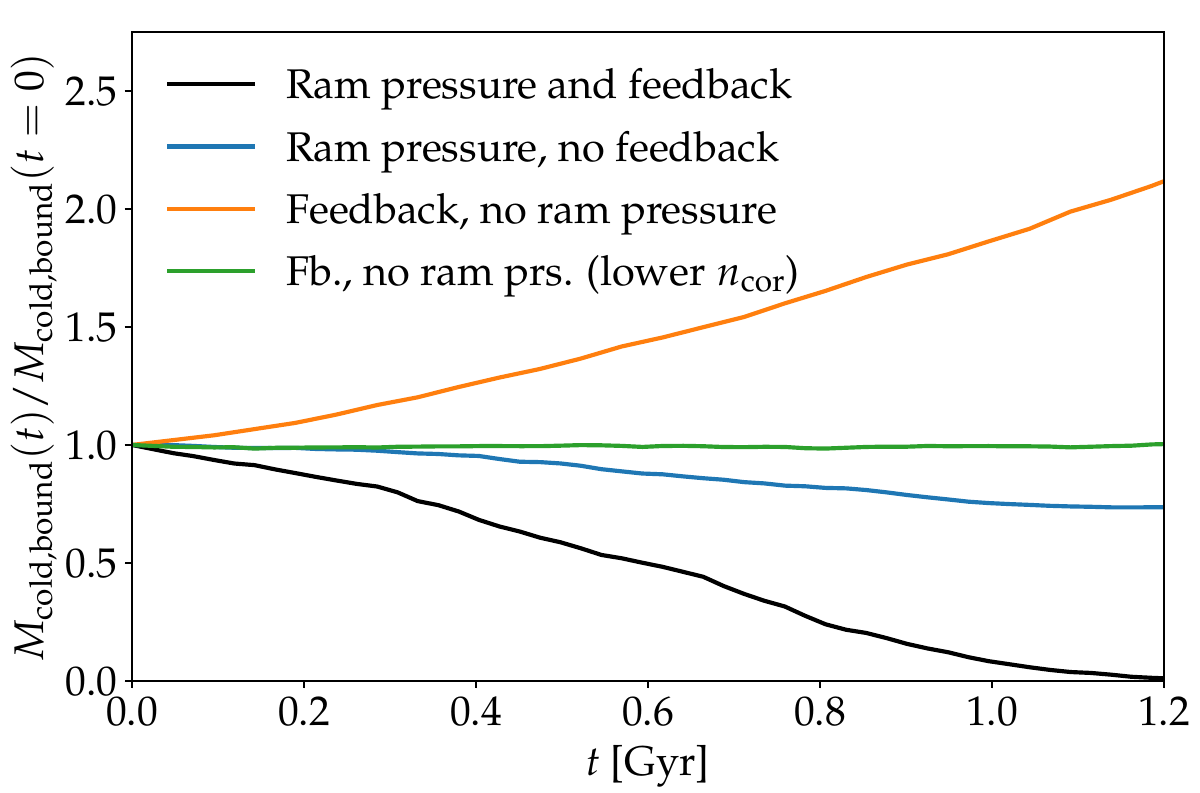}
    \caption{Evolution of the mass of cold bound gas for our fiducial simulation compared to simulations without feedback or without ram pressure. All simulations have $n_{\mathrm{cor}}=8\times 10^{-4}$ cm$^{-3}$ except the green curve which has $n_{\mathrm{cor}}=10^{-4}$ cm$^{-3}$.}
    \label{fig:fbvsnofb}
\end{figure}

To assess how the SN feedback affects stripping we run variations of our fiducial run. In one case, we switch off the feedback and in the other two cases we do not inject a wind such that there is no ram pressure. We show the evolution of the cold bound gas mass in these simulations compared to the fiducal run in Figure \ref{fig:fbvsnofb}. Clearly, without feedback the ram pressure stripping proceeds much more slowly: $M_{\mathrm{cold, bound}}$ is still 73 per cent of its initial value at the end of the simulation with $n_{\mathrm{cor}}=8\times 10^{-4}$ cm$^{-3}$ that leads to complete stripping with feedback.

On the other hand, feedback by itself is unable to remove any significant mass of ISM. In fact, at $n_{\mathrm{cor}}=8\times 10^{-4}$ cm$^{-3}$ we find that the mass \emph{increases} substantially with time in the simulation with no injected wind. In this case, the mixing between the ejected ISM and the relatively dense hot gas at the ISM-corona boundary leads to condensation. That is, coronal gas cools and accretes onto Draco in a process analogous to accretion through Galactic fountain condensation near the disc-corona interface of the MW \citep{marinacci10,gronnow18}. We do not suggest, though, that Draco, or any other MW satellite, is actually growing its ISM through fountain-like accretion. This process is easily disrupted by ram pressure which removes ejected gas, as seen in our other simulations, preventing it from falling back onto the ISM and bring in additional condensed gas. At a lower coronal density of $n_{\mathrm{cor}}=10^{-4}$ cm$^{-4}$, more representative of the present day corona, this accretion process is also not effective even without ram pressure. Instead, $M_{\mathrm{cold, bound}}$ oscillates very slightly but is overall essentially conserved as ejected gas is unable to escape but also unable to effectively cool the surrounding hot gas.

Our finding that SN feedback greatly hastens the ram pressure stripping agrees with most previous studies \cite[e.g.][]{gatto13,bahe15,samuel22}. It is in contradiction of \cite{emerick16} who found that specifically for a low mass $M_\star \sim 10^5 M_\odot$ galaxy, the inclusion of feedback had little effect on the stripping. They had to artificially boost the SN rate by a factor of $\approx 5$ for feedback to become important. However, their simulated galaxies had a factor of $3-4$ lower initial SFRs, and thus correspondingly lower SN rates, than Draco and a further factor of $\approx 2$ lower SN rate due to them adopting a Salpeter IMF while we adopt a \cite{chabrier03} IMF. With that in mind, our result agrees with their result that feedback did become important for ram pressure stripping with factor $\approx 5$ higher SN rates. Our result that feedback alone is inefficient at removing the ISM in low mass dwarfs like Draco also agrees with previous studies \citep[e.g.][]{caproni17,bermejo-climent18,donatella19}.

\subsection{Lower and higher virial mass}
\label{sec:mvir}
\begin{figure}
    \centering
    \includegraphics[width=0.49\textwidth]{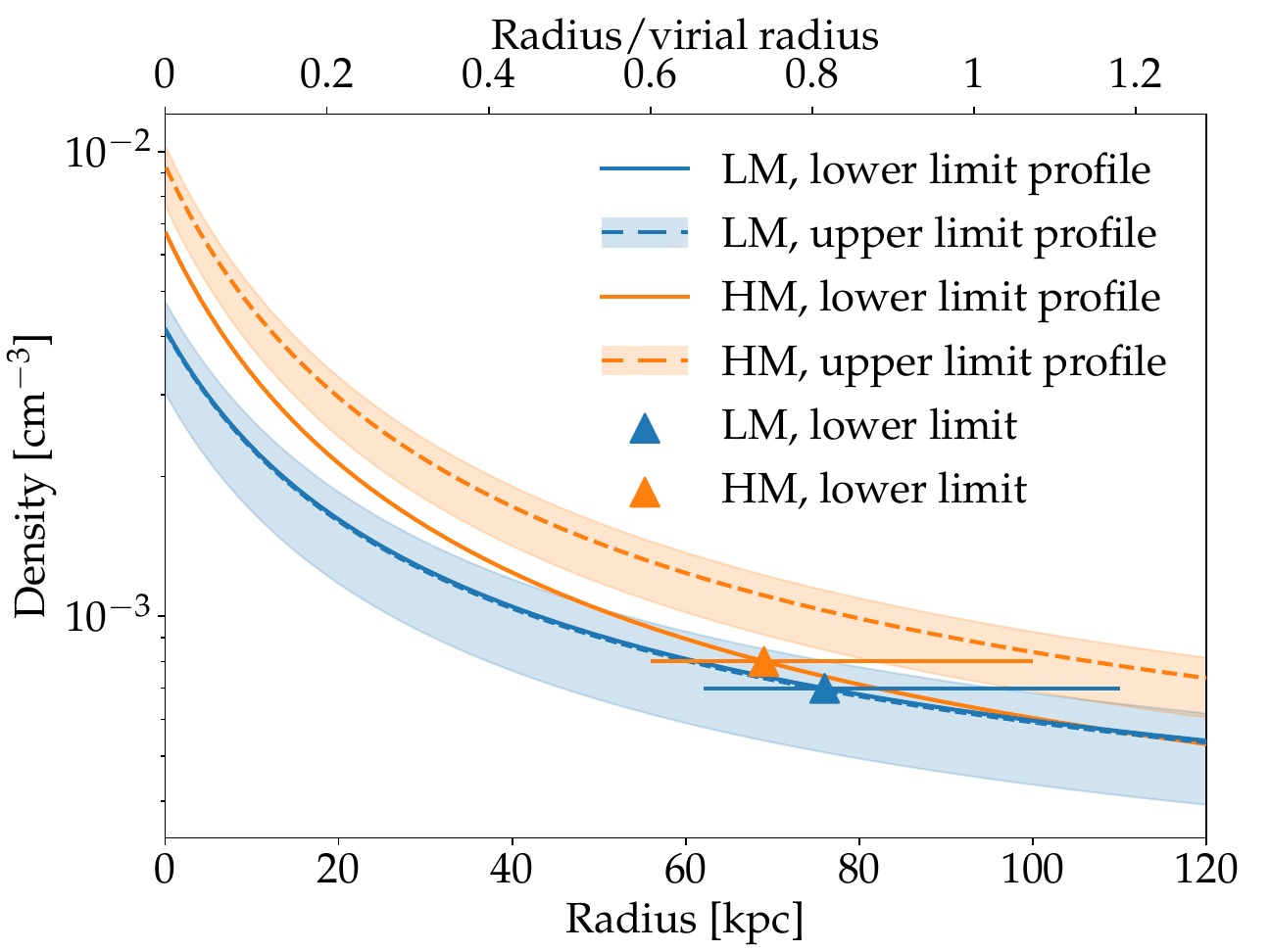}
    \caption{Similar to Figure \ref{fig:densprofz16}, but for the lower $9.5\times 10^{11} M_\odot$ (`LM', blue) and higher $1.6\times 10^{12} M_\odot$ (`HM', orange) MW virial mass.}
    \label{fig:densprofz16-lowhigh}
\end{figure}

So far we have exclusively considered our fiducial choice for the present day MW virial mass of $M_{\mathrm{vir, z=0}}=1.25 \times 10^{12} M_\odot$. As mentioned in Section \ref{sec:orbitint}, two other less likely, but still possible, virial masses of $M_{\mathrm{vir, z=0}}=9.5\times 10^{11} M_\odot$ and $M_{\mathrm{vir, z=0}}=1.6\times 10^{12} M_\odot$ also lead to a first passage that is consistent with ram pressure stripping. At the lower/higher mass, Draco has completed one less/more passage at the present day. We have run simulations varying the coronal density for these orbits as well. The behaviour is rather similar with complete stripping occurring at $\langle n_{\mathrm{cor}}(62-110$ kpc$)\rangle=7\times 10^{-4}$ cm$^{-4}$ for the lower mass and at $\langle n_{\mathrm{cor}}(56-100$ kpc$)\rangle=8\times 10^{-4}$ cm$^{-4}$ for the higher mass. Thus, the effect on the stripping due to slightly lower/higher pericentric velocity at the lower/higher mass appears to be largely cancelled out by the longer/shorter orbital period (we run the lower and higher mass simulations for 1.4 Gyr and 1.1 Gyr, respectively). We show the lower and upper limits on the density at the low and high mass in Figure \ref{fig:densprofz16-lowhigh}.

At $M_{\mathrm{vir, z=0}}=1.6\times 10^{12} M_\odot$ there is the slight complication that the injected velocity becomes supersonic for a relatively small part of the orbit around the pericentre. This is a consequence of our injection velocity being higher than the actual orbital velocity as described in Section \ref{sec:velinj}. The actual orbital velocity comes close to, but never exceeds, the sound speed. Hence the resulting shock is artificial. Because the injection velocity only ever becomes slightly supersonic we can approximate the orbit quite well by restricting the injection velocity near pericentre to be slightly below the sound speed. Fortunately, the stripping in this simulation agrees very closely with the simulation without the restricted injection velocity where the shock near pericentre is allowed to occur. This indicates that the relatively weak shock is of little consequence.

While the inferred lower limit on the coronal density along the orbit barely differs for the three masses, the derived isothermal profiles differ more substantially. These changes are due to the different masses and concentrations of the halo (the virial radius at the time of pericentre is approximately equal for all three masses, however). This affects both the shapes of the lower limit profiles and the inferred upper limit profiles due to changes in the allowed baryonic mass. As can be seen, the difference between the two density profiles is greatest at the centre where the density differs by about 60 per cent. Additionally, the pericentric passage occurs at slightly different times as shown in Table \ref{tab:orbits}. The derived constraints are shown, together with the fiducial virial mass case, in Table \ref{tab:results}. For the low mass, $n_{\mathrm{cor,min}}=7\times 10^{-4}$ cm$^{-4}$ is approximately equal to the upper limit. For the high mass, $n_{\mathrm{cor,min}}=8\times 10^{-4}$ cm$^{-4}$ is still close to the upper limit, but farther below than the other cases.

Extrapolating the coronal masses of the low and high virial masses to the present day yields quite similar results as for the fiducial virial mass. As expected, the upper limit on $f_\mathrm{cor}$ (i.e. when no gas is ejected from the corona) is slightly lower/higher for the lower/higher virial mass. For the low mass $f_\mathrm{cor,max}=0.71$ while for the high mass $f_\mathrm{cor,max}=0.85$.

\subsection{Lower initial SFR}
\label{sec:lowsfr}
Draco's initial ISM mass is based on the SFR in the bin centred at a cosmic age of $t\approx 3.5$ Gyr in the SFH (see Section \ref{sec:sfh}). This SFR has an uncertainty of $\sigma\approx 1.2 \times 10^{-5} M_\odot$ yr$^{-1}$ around the $5.8\times 10^{-5} M_\odot$ yr$^{-1}$ value that we generally adopt. We expect that higher initial SFR generally entails higher required coronal densities for complete stripping, $n_{\mathrm{cor,min}}$, as the initial ISM mass will be higher, although the additional disruption from the higher SN rate will partially counteract this. Due to $n_{\mathrm{cor,min}}$ at our standard SFR already being close to the upper limit from the cosmological baryon fraction, we focus on the evolution in the case of 1 $\sigma$ lower initial SFR, i.e. $\Psi_{\mathrm{tot}} = 4.6\times 10^{-5} M_\odot$ yr$^{-1}$.

We find that in this case $n_{\mathrm{cor,min}}$ is indeed lower than for the fiducial initial SFR, although only just slightly, at $n_{\mathrm{cor,min}}=7\times 10^{-4}$ cm$^{-3}$. This difference is approximately the same as the 15 per cent difference in the initial ISM mass. Hence, the uncertainty in Draco's initial SFR from the SFH does not change our results substantially.

\subsection{Coronal temperature}
\label{sec:cortemp}
We have assumed a temperature of $T_\mathrm{cor}=2.2 \times 10^6$ K for the corona following the median temperature found by \cite{henley13}. They find fairly uniform temperatures with variations along different lines of sight of $\pm 6.3 \times 10^5$ K around this median. This is in quite good agreement with other studies \citep{yoshino09,gupta12,gupta21} although \cite{nakashima18} found a higher temperature of $\approx 3 \times 10^6$ K. A lower/higher coronal temperature leads to a higher/lower initial ISM mass due to the initial pressure equilibrium. Additionally, a hotter corona can more effectively heat stripped gas through mixing, although heating of stripped gas is already very efficient in our simulations and this gas is unimportant for our analysis because it would become unbound from Draco anyway.

\begin{figure}
    \centering
    \includegraphics[width=0.49\textwidth]{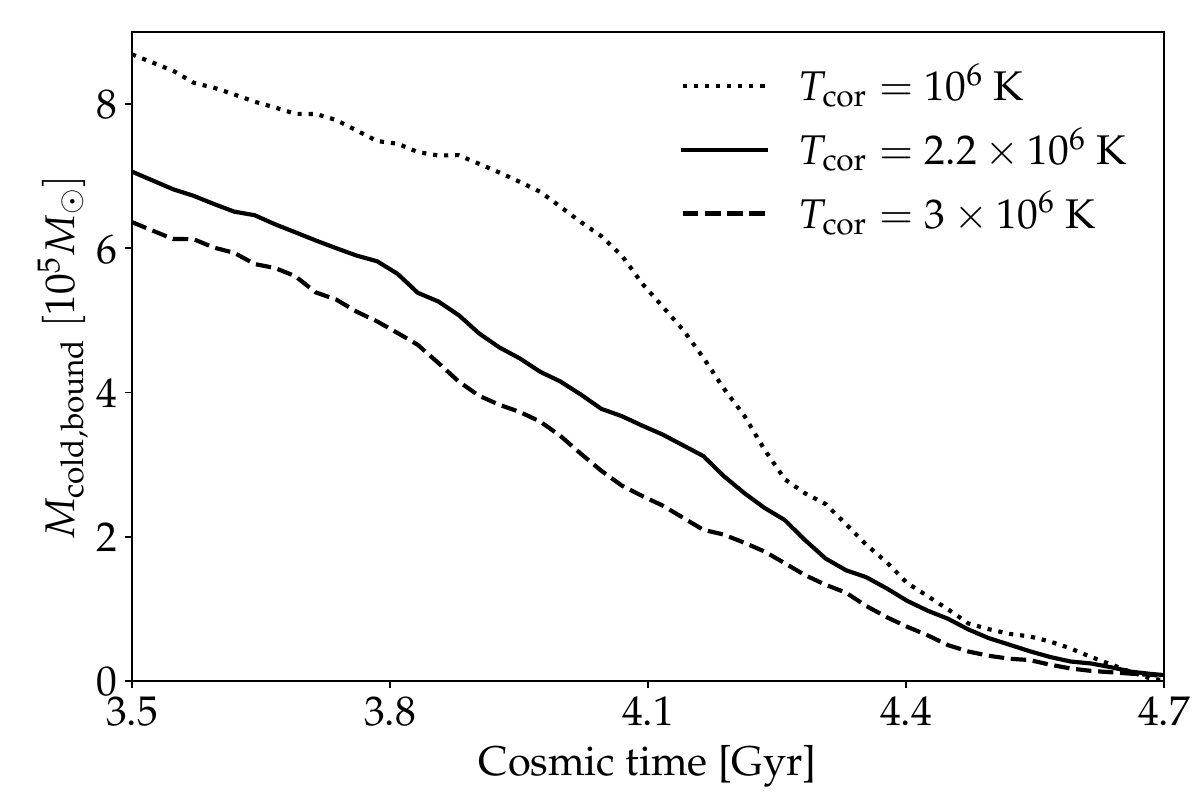}
    \caption{Evolution of the mass of cold bound gas for our fiducial simulation at $n_{\mathrm{cor}}=8\times 10^{-4}$ cm$^{-3}$ (solid line) compared to simulations with the same coronal density but lower (dotted line) and higher (dashed line) coronal temperatures.}
    \label{fig:temperature}
\end{figure}

To assess the effect of changing our assumed coronal temperature, we have run the simulation with $n_{\mathrm{cor}}=8\times 10^{-4}$ cm$^{-3}$, that we found to be the minimum coronal density required for complete stripping, at lower and higher $T_\mathrm{cor}$. These are approximately 50 per cent lower and higher than the fiducial temperature at $10^6$ K and $3\times 10^6$ K, respectively. We compare the evolution of the mass of cool bound gas at the low, fiducial, and high temperatures in Figure \ref{fig:temperature}.

While the initial ISM mass decreases with increasing coronal temperature, the ISM is lost at a correspondingly lower rate around the pericentric passage such that complete stripping occurs at essentially the same time in all three cases. In general, we would also not expect this additional initial ISM to have much impact. Because the density profile is unchanged, all the additional ISM at lower coronal temperatures is due to its larger initial extent. This gas, while occupying a large volume and so contributing a decent fraction to the mass, is at low densities and can be stripped very effectively.

\cite{gatto13} found that a significantly lower coronal temperature led to a significantly higher $n_{\mathrm{cor,min}}$. However, a significantly higher temperature only led to a slightly lower $n_{\mathrm{cor,min}}$. While we do not find significant differences in either case, we do find the same relative trend. That is, when normalizing the mass by $M_{\mathrm{cold, bound}}(t=0)$ a slight difference remains during the middle part of the simulations with the lower(higher) coronal temperature simulation having more(less) cold bound gas compared to the fiducial case. For example, while half of the ISM has been lost by $t_{\mathrm{peri}}$ at the fiducial temperature, 63(40) per cent has been lost at the lower(higher) temperature. Still, the clear conclusion is that the choice of coronal temperature in our simulations has no significant effect within the plausible temperature range.

The change in coronal temperature does alter the shape of our derived isothermal density profiles, which become steeper at lower temperature. However, the total coronal mass does not change significantly for the lower temperature, which still has a lower limit baryon fraction of $f_{\mathrm{b,min}}=0.17$. At the higher temperature the corona is slightly, but not significantly, too massive compared to the upper limit from the cosmological baryon with a minimum baryon fraction of $f_{\mathrm{b,min}}=0.19$. Our conclusions from the extrapolation to the present day in Section \ref{sec:densprofz0} likewise hold at the lower and higher temperatures. Hence, our finding that the corona could contain all the baryons expected from the cosmological baryon fraction is robust to changes in the coronal temperature.

\subsection{Density variation along the orbit}
\label{sec:vardens}
So far we have used a constant coronal density for all simulations, while varying the velocity along the orbit. We then interpreted this density as the average density of the corona along the simulated part of the orbit. This hinges on the assumption that the orbital velocity dominates the ram pressure due to the ram pressure depending linearly on the coronal density but on the square of the velocity. To assess the validity of this, we have run a simulation where we also vary the density with time according to the density profile of Figure \ref{fig:densprofz16}. Hence, the average weighted density along the orbit $\langle n_{\mathrm{cor}} \rangle=8 \times 10^{-4}$ cm$^{-3}$ matches that of the minimum constant density that we found to be required for complete stripping.

In order to have the corona still be stable within our setup that does not explicitly include the MW potential we have to keep the gas pressure constant. Hence, we have to vary the temperature with the density along the orbit such that it decreases (increases) as Draco moves towards (away from) the pericentre. While this is inconsistent with the density profile being isothermal, this temperature variation is not an issue in practice as we have shown in Section \ref{sec:cortemp} that changing the coronal temperature by a factor two has a negligible effect on the stripping.

\begin{figure}
    \centering
    \includegraphics[width=0.49\textwidth]{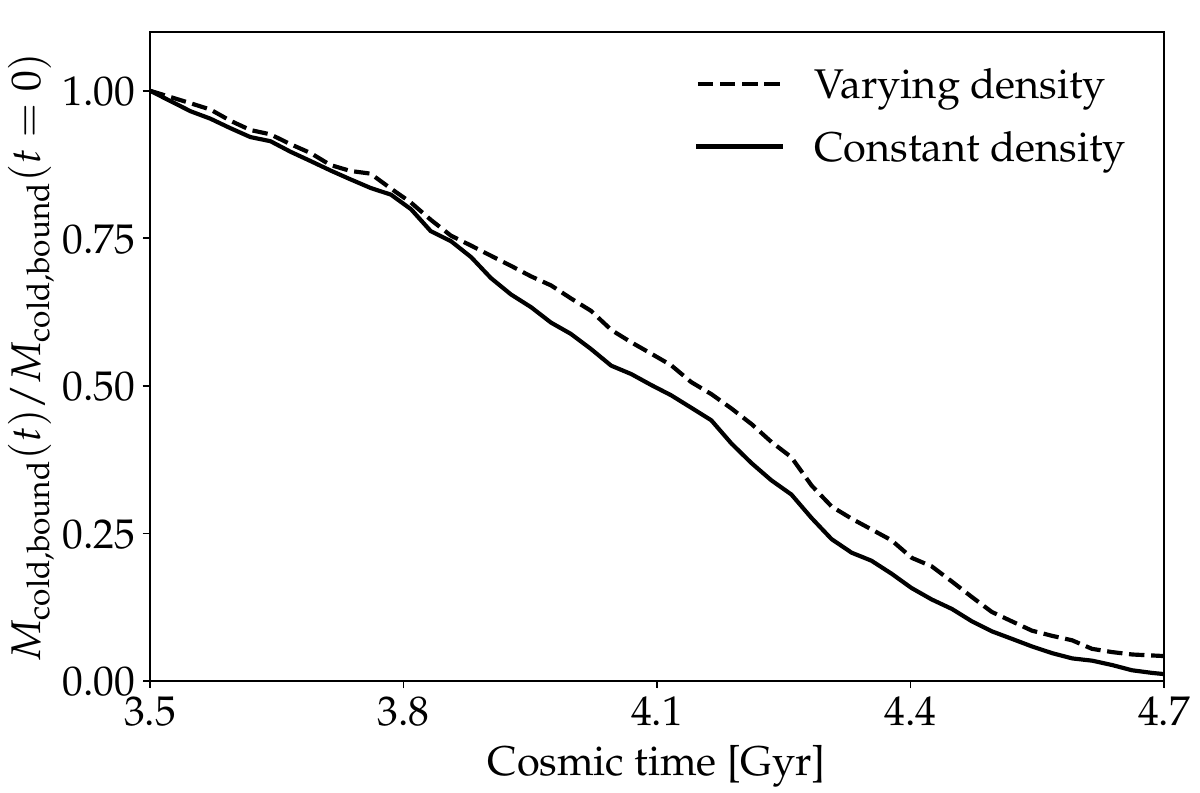}
    \caption{Evolution of the mass of cold bound gas for our fiducial simulation with a constant coronal density of $n_{\mathrm{cor}}=8\times 10^{-4}$ cm$^{-3}$ (solid line) compared to a simulation with a varying coronal density with the same average density (dashed line).}
    \label{fig:vardens}
\end{figure}

We compare the evolution of the mass of cool bound gas for the varying and constant coronal density simulations with $\langle n_{\mathrm{cor}} \rangle=8 \times 10^{-4}$ cm$^{-3}$ in Figure \ref{fig:vardens}. As can be seen, the differences in the mass evolution are negligible, demonstrating that a constant coronal density is a valid approximation.

\subsection{Resolution}
\begin{figure}
    \centering
    \includegraphics[width=0.49\textwidth]{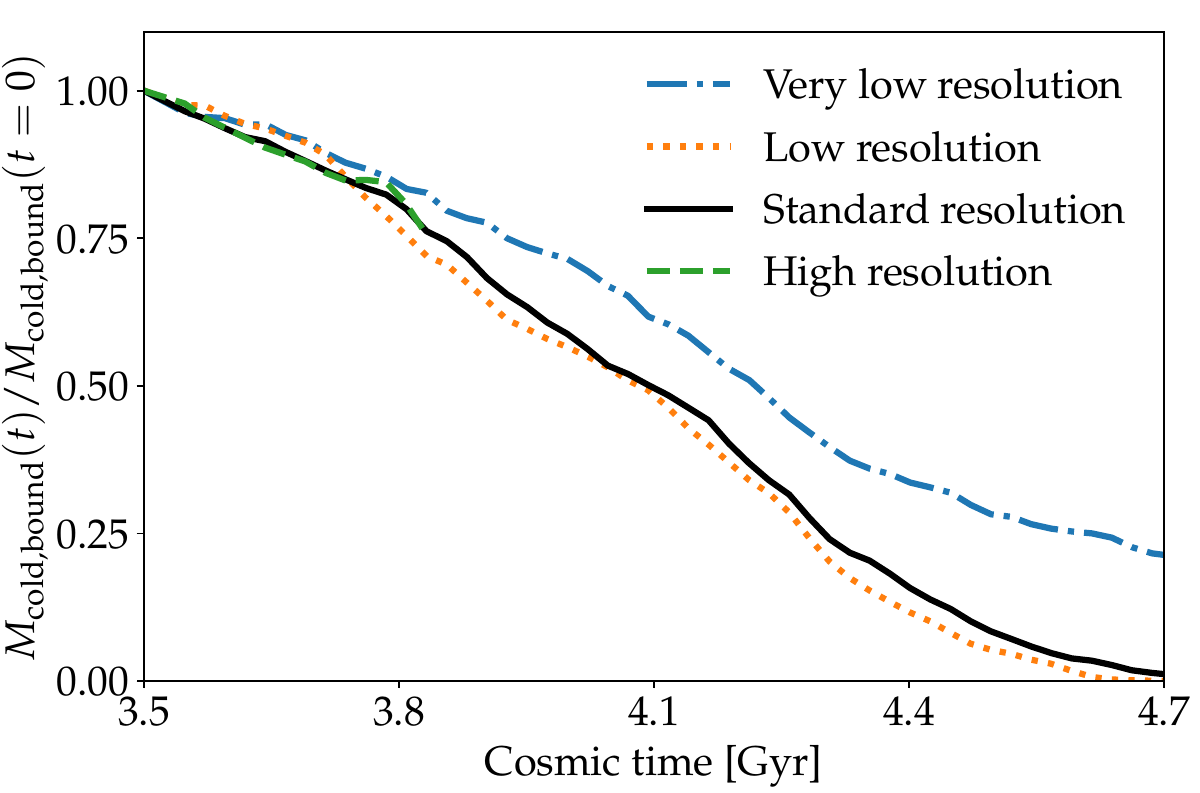}
    \caption{Evolution of the mass of cold bound gas for our fiducial simulation at $n_{\mathrm{cor}}=8\times 10^{-4}$ cm$^{-3}$ (black solid line) compared to simulations with one quarter (blue dot-dashed line), half (orange dotted line), and twice (green dashed line) the fiducial maximum resolution.}
    \label{fig:resolution}
\end{figure}
We run simulations similar to our fiducial with a coronal density of $n_{\mathrm{cor}}=8\times 10^{-4}$ cm$^{-3}$ but with different resolutions to assess its effect on the stripping. These simulations have the same simulation domain but with a different number of AMR levels leading to lower or higher maximum resolutions. We run a `very low' resolution simulation with up to level $n=9$, a `low resolution' simulation with up to level $n=10$ and a `high resolution' simulation with up to level $n=12$. This corresponds to one quarter, half, and twice the standard resolution, respectively, in terms of the maximum number of cells per kpc. As is the case at our standard resolution, maximum refinement is always enforced within 400 pc of the centre of Draco's potential and is otherwise refined according to mass following the criterion given in Section \ref{sec:nummethods}. This ensures that essentially all the cold gas is at the highest allowed resolution in each simulation. The low-density wake behind Draco consequently remains on a low refinement level, and hence unresolved, also in the high resolution case. However, this unbound gas is in any case not important for our purposes.

We show the evolution in the mass of cold bound gas, $M_{\mathrm{cold, bound}}$, in Figure \ref{fig:resolution}. As can be seen, there is little difference between the low, standard, and high resolution cases. Due to the computational cost and challenging numerical stability of the high resolution simulation we only ran this until about $t=3.8$ Gyr, about one quarter of the time of the other simulations. However, within this time span the evolution is clearly essentially the same as at the standard resolution. The high resolution simulation does have a more prominent tail leading to a bit ($\approx$10 per cent) higher overall mass of cold gas, but this is not relevant for to our analysis because this gas is not bound to Draco. While $M_{\mathrm{cold, bound}}$ in the low resolution case is not significantly different from the standard resolution case, it does differ slightly more than the high resolution case, indicating that the evolution of $M_{\mathrm{cold, bound}}$ has largely converged at our standard resolution. This shows that our feedback scheme is indeed well resolved already at the $\Delta x\approx 23$ pc resolution of the low resolution simulation. Meanwhile, as expected, $M_{\mathrm{cold, bound}}$ is far from converged at the very low resolution, with this simulation considerably underestimating the stripping. Presumably, this is due to the feedback not being resolved in this case such that the energy injected by SNe does not couple as effectively to the surrounding gas as it should.

\subsection{Gas distributions}
\label{sec:gasdist}
We assume isothermal profiles for Draco's ISM and for the MW corona. In reality, Draco must initially have contained some colder, denser molecular gas. While we do allow cooling down to a temperature of $\approx 150$ K and include partial self-shielding, we do not include molecular cooling which should be particularly important at the low metallicity of Draco's ISM \citep{glover14}. In general, UV heating still prevents any significant amounts of gas to cool below $\approx 6000$ K in our simulations due to our partial suppression of self-shielding (see Section \ref{sec:nummethods}). \cite{hausammann19} simulated satellite galaxies extracted from zoom-in simulations that included cold gas at $T\approx 10$ K. They found that while ram pressure easily stripped the warmer $T>1000$ K gas, it was inefficient at removing the remaining core of cold gas. Their assumed CGM density was relatively low at $\sim 10^{-5}$ cm$^{-3}$ and their dwarfs a factor of $\gtrsim 3$ more massive than Draco, which would lead to generally less effective ram pressure stripping than in our simulations. In any case though, more realistic ISM distributions and cooling models that allow for cold clumps should be investigated in future studies aimed at putting constraints on the coronal density from ram pressure.

In our simulations, we use constant densities for the corona. Because of this, we do not assume a specific density profile to derive our lower limit on the coronal density, at the cost of then only constraining the average density along the orbit. We do assume isothermal density profiles in hydrostatic equilibrium with the MW NFW potential for extrapolating our lower limit to other radii and to the present day and to derive upper limits. These profiles are somewhat flatter than most profiles used in emission and absorption studies. These often fit a (essentially) two parameter $\beta$-profile where the slope is given by $\beta$ and outside of the innermost parts is well approximated by a simple power law $n\propto r^{-3\beta/2}$ \citep[e.g.][]{miller15,li17,bregman18}. This model is generally not isothermal, instead the temperature required for hydrostatic equilibrium tends to decrease with radius \citep{guo20}. However, due to the density dependence of these observations they only constrain the density at $r \lesssim 50$ kpc \citep{bregman18}. Our isothermal profiles for the present day corona (see Section \ref{sec:presentcorona}) are well approximated outside of the innermost parts at $r<10$ kpc (where, as previously mentioned, we would expect our profile to be too flat due to our ignoring the disc potential) by a $\beta$-profile with $\beta=0.3$. This is on the lower side of estimates that typically find $\beta \approx 0.5$ \citep{miller13,miller15} but in agreement with absorption constraints if a metallicity gradient is included \citep{bregman18} and preferred by some models \citep{faerman17,faerman22,martynenko22}. Additionally, X-ray emission suggests that the hot gas in the innermost part of the halo is dominated by a disc-shaped component \citep{yao09,nakashima18} and models that include such a structure can accommodate a flatter profile for the larger scale spherical CGM component \citep{kaaret20,yamasaki20}.

In any case, a spherically symmetric density distribution can be an approximation. \cite{simons20} examined the ram pressure stripping of satellites radially infalling in MW-like haloes in the FOGGIE cosmological zoom in simulations. These haloes have highly locally structured CGM and consequently the ram pressure stripping mainly occurs stochastically and suddenly when a satellite collides with a denser filament, rather than from the hot gas. However, this analysis was done at $z\geq 2$ when the simulated galaxies had not yet formed significant hot coronae. On the contrary, the hot CGM in simulated galaxies appears to be remarkably spherical even in the presence of strong outflow given the high degree of mixing \cite[e.g.][]{Gutcke+2017}.

Likewise, our assumed spherically symmetric DM, stellar, and initial ISM profiles for Draco are only approximations. \cite{hayashi20} found Draco's DM profile to be consistent with a spherical profile within 1$\sigma$, however their large uncertainties in axis ratio also allowed for it to be quite non-spherical. In any case, with an apparent ellipticity of $\epsilon=0.3$ in Draco's present day stellar distribution \citep{munoz18}, it is clearly not spherical but also not highly elongated. Modelling Draco using ellipsoidal density distributions would introduce a dependence on the angle between its major axis and its trajectory. However, we expect that this dependence would be relatively weak based on Draco's relatively circular apparent shape.

\subsection{Milky Way potential}
The hydrostatic density profile for the corona, the coronal mass growth, and, in particular, the orbit of Draco depend on our assumed model for the MW's potential, for which we have chosen \cite{zhao09}. Fortunately, other studies of the growth of haloes based on different N-body simulations in the literature agree well on the mass accretion history, and hence the potential, in the parameter space relevant for our study \citep[i.e. MW progenitor mass haloes at redshifts $z<2$ in a standard $\Lambda$CDM cosmology, e.g.][]{fakhouri10,vandenbosch14,correa15}. We have confirmed that these models yield similar virial masses and radii and orbits of Draco. \cite{buist14} developed a model for the growth of the potential that differs more substantially from \cite{zhao09}. In order to more correctly capture the inside out growth of haloes, this model ensures that the halo density at fixed radius always increases. Consequently, it leads to a stronger potential at early times which causes Draco to have an additional earlier passage at $z\approx 4$. However, the pericentre of this passage is more than 50 kpc from the Galactic centre, which is much further than the virial radius at this very early time. Hence, it is unlikely that the corona, if it had even formed yet, would have extended sufficiently far to affect Draco. Hence, the earliest passage where stripping from the corona is plausible would still be the subsequent passage at $z<2$. At that point, the evolution is largely similar to that of the \cite{zhao09} model and hence this second passage does not differ substantially from the first passage that we simulate.

\cite{gaia18} found that Draco and Ursa Minor have quite similar orbits and suggested that they could have fallen in to the MW halo as part of a group. In this case, additional disruption caused by the gravitational interaction with Ursa Minor could lead to a lower coronal density required for stripping.

\subsection{Coronal rotation}
In this work, we have assumed that the corona is at rest in the reference frame of the dark matter halo (equivalently, that the relative velocity between Draco and the corona is the same as the orbital velocity of Draco). This assumption is probably not exactly correct, as the corona itself is also expected to have some internal motions (e.g. \citealt{oppenheimer18}) and certainly some amount of rotation, as it must have non-negligible angular momentum (e.g.\ \citealt{teklu15}; \citealt{pezzulli17}; \citealt{Sormani+2018}).

Unfortunately, the exact amount of rotation is not known empirically, as observational constraints on the coronal kinematics are challenging at $z = 0$ (\citealt{hodges-kluck16}) and non-existent at $z = 1.6$.  We can however discuss the possible impact of coronal rotation in light of theoretical considerations. We focus in particular on the pericentric passage, motivated by the considerations that (i) this is the most crucial moment for stripping and (ii) by coincidence, the orbit of Draco is such that at the moment of pericentric passage its velocity is almost entirely in the direction of rotation ($v_\textrm{peri} \sim v_\textrm{rot,peri} = 210 \; \textrm{km} \; \textrm{s}^{-1}$) and hence the effect of the rotation of the corona could in principle affect the results in a non-negligible way.

Because the ram pressure force is proportional to $(\Delta v)^2$, its magnitude would, for instance, be halvened (with an arguably important effect on our results) if the corona was rotating at $v_\textrm{cor} = ((\sqrt{2}- 1)/\sqrt{2}) v_\textrm{Draco} = 61.5 \; \textrm{km} \; \textrm{s}^{-1}$.
The pericentre in cylindrical coordinates is ($R_\mathrm{peri} = 20$ kpc$, \vert z_\mathrm{peri}\vert=56$ kpc). This cylindrical distance corresponds to an angular momentum of 1230 km $\textrm{s}^{-1}$ kpc, or about twice the expected average angular momentum of the dark matter at that redshift (653 km $\textrm{s}^{-1}$ kpc, following e.g.\ \citealt{bullock01}). Although the corona is also expected to contain gas with specific angular momentum exceeding the dark matter average (e.g.\ \citealt{pezzulli17}), models show that this high angular momentum gas should be primarily located at large cylindrical radii and relatively close to the equatorial plane (in the regime $R > \vert z\vert$ using standard cylindrical coordinates), as shown, for instance, in the Appendix of \cite{afruni22}. This is the opposite regime as the one relevant to the almost polar orbit of Draco whose orbital plane is inclined by only 20 degrees from the rotation axis of the Galaxy. Hence, a particularly large angular momentum of the corona, capable of substantially altering our results, is not expected in the region of interest for our study. A more precise estimate of the rotation velocity of the corona in the region of our interest is beyond the scope of this work and also unavoidably model dependent. However, we do notice that reading the model of \cite{afruni22} at the pericentric position of Draco (after re-scaling by the different value of the virial radius) would result in a coronal angular momentum of about 50 per cent of the DM average. This would correspond to a mere 15 km $\textrm{s}^{-1}$ of coronal rotation velocity and a modest 15 per cent impact on the ram pressure force at pericentre. Therefore, although there is undoubtedly some degree of uncertainty in this kind of estimate, we conclude that it is unlikely that the rotation of the corona could have a large impact on our results.

\section{Conclusions}
\label{sec:conclusions}
Using the SFH and orbit of Draco, taking the evolution of the MW halo into account, we have shown that it was likely ram pressure stripped by the early corona during its first infall. By simulating this passage covering redshifts $z\approx 1.3-1.9$ we have derived a lower limit on the average density in the outer part of the early corona relatively shortly after it formed. For all three allowed present day virial masses we find $n_{\mathrm{cor,min}}\approx 7-8 \times 10^{-4}$ cm$^{-3}$ at about $3/4$ of the virial radius at that time. We find that this does not depend significantly on the assumed coronal temperature and is well converged at our standard resolution. The simple analytical estimate (eq.~\eqref{eq:ramstrip}) turns out to be in excellent agreement with this, although we consider this to be a coincidence because the assumption of instantaneous stripping is not satisfied. We find that the inclusion of SN feedback is crucial to the efficiency of ram pressure stripping. However, by itself it is inefficient at removing the ISM.

We extrapolated our lower limit density to other radii within the virial radius assuming an isothermal profile that is in hydrostatic equilibrium in the MW's NFW potential at the time of pericentre ($z=1.6$ at our fiducial virial mass). This leads to coronal masses that are allowed within the cosmological baryon fraction but would have contained almost all the `missing' baryons. Extrapolating this mass to the present day assuming a constant fraction $f_{\mathrm{cor}}$ of baryons accreted onto the halo end up in the hot corona, we showed that this leads to densities that are in decent agreement with the literature constraints. The present day coronal masses are also below the cosmological baryon fraction for all values of $f_{\mathrm{cor}}$ below the upper limit of 0.8 given by the growth of the disc. In any case, the overall average density of the corona has decreased substantially, probably by an order of magnitude. Most of this evolution presumably occurred at $z>1$ when the accretion rate onto the halo was higher and most of the growth in the virial mass and radius happened as well. Studies of ram pressure stripping should take this into account if they are considering satellites that could have been stripped early.

Due to the massive improvement over previously available proper motion measurements by \emph{Gaia}, the limiting factor for ram pressure stripping studies like \cite{gatto13} and our work has instead become the scarcity of SFHs in the literature. For the present day MW corona, the Fornax \citep{deboer12a,rusakov20} and Carina \citep{deboer14,savino15,santana16} satellite galaxies have detailed SFHs available showing that they lost their gas during a recent passage. For probing the early corona, two potential targets other than Draco are the Ursa Minor \citep{carrera02,dolphin03} and Sculptor \citep{deboer12b,savino18,betinelli19} satellite galaxies. Star formation ended early in both of these dwarfs, but not so early as to be caused by reionisation. Interestingly, the estimated coronal densities required to strip them from eq. \eqref{eq:ramstrip} are very high, with Sculptor in particular requiring densities generally above our estimated upper limit (see Section \ref{sec:presentcorona}). This also indicates that they were stripped in the earlier more dense corona. Future ram pressure simulations could target these to further investigate the early MW corona.

\section*{Acknowledgements}
AG, FF and GP acknowledge support from the Netherlands Research School for Astronomy (Nederlandse Onderzoekschool voor Astronomie, NOVA), Network 1, Projects 10.1.5.7 (AG and FF) and 10.1.5.18 (GP).

We acknowledge PRACE for awarding access to the Fenix Infrastructure resources, which are partially funded from the European Union’s Horizon 2020 research and innovation programme through the ICEI project under the grant agreement No. 800858.

We thank SURF (www.surf.nl) for the support in using the Dutch National Supercomputer Snellius.

We acknowledge the CINECA award (project no. IsC78\_RAMPRESS) under the ISCRA initiative, for the availability of high performance computing resources and support.

AG and FF would like to thank the Center for Information Technology of the University of Groningen for their support and for providing access to the Peregrine high performance computing cluster.

\section*{Data availability}
The code that we used to generate our simulations is available at \url{https://github.com/agronnow/ram_pressure_stripping}. Simulation output data will be shared on reasonable request to the corresponding author.




\bibliographystyle{mnras}
\bibliography{rampressurestripping}


\appendix

\section{Growth of the corona}
\label{sec:coronaevol}
The initial corona at redshift $z\gtrsim 2$ would have been much denser and more compact than today as it has since expanded as the virial radius of the MW has grown. While the density is expected to have decreased, the total mass of the corona has grown through infall of gas from the IGM that is unable to reach the disc as well as hot gas expelled from the MW's ISM by supernovae.

We refer to all the gas within the virial radius but outside of the ISM as `circumgalactic medium' (CGM). From mass conservation it then follows that the mass in the CGM must change according to the sum of four non-negative terms:
\begin{equation}
\label{eq:dmcordt}
\dot{M}_{\mathrm{CGM}}=\dot{M}_{\mathrm{IGM\rightarrow CGM}}+\dot{M}_{\mathrm{ISM\rightarrow CGM}}-\dot{M}_{\mathrm{CGM\rightarrow IGM}}-\dot{M}_{\mathrm{CGM\rightarrow ISM}}.
\end{equation}
The arrows indicate the direction of the mass flow. These terms represent inflow of gas from the IGM, gas ejected from the ISM, gas ejected from the CGM, and gas accreted onto the ISM, respectively.

While some of the gas in the CGM at any point in time will be cold, any cold cloud or filament cannot remain in the CGM. It will either be heated or fall onto the disc. Gas that remained cold throughout its journey onto the disc is commonly referred to as `cold-mode accretion' while gas that was heated and then later cooled to eventually make its way to the disc is referred to as `hot-mode accretion' \citep[see e.g.][]{birnboim03,keres05}. Thus, cold gas will not accumulate in the CGM over longer time-scales. This is supported also by the results of cosmological zoom-in simulations which find that the fraction of the CGM mass that is in hot gas increases substantially with time \citep{hafen19}. Considering then the change in mass over a time $\Delta t$ much longer than the time-scale for cold gas infall (which is estimated to be of the order of $\lesssim 2$ Gyr at $z\leq 2$; see \cite{nelson15}) we can equate the change in the CGM mass over that time with the change in the hot corona mass $\Delta M_{\mathrm{cor}}$.
Hence, from eq. \eqref{eq:dmcordt} we can derive
\begin{equation}
\label{eq:deltamcor}
\Delta M_{\mathrm{cor}}=f_\mathrm{cor}\Delta M_{\mathrm{IGM\rightarrow CGM}},
\end{equation}
where
\begin{equation}
f_{\mathrm{cor}}\equiv1-\frac{\Delta M_{\mathrm{CGM\rightarrow ISM}} + \Delta M_{\mathrm{CGM\rightarrow IGM}} - \Delta M_{\mathrm{ISM\rightarrow CGM}}}{\Delta M_{\mathrm{IGM\rightarrow CGM}}}.
\end{equation}
$f_\mathrm{cor}<0$ would imply that the corona is losing mass while $f_\mathrm{cor}>1$ would imply that $\Delta M_{\mathrm{ISM\rightarrow CGM}} > \Delta M_{\mathrm{CGM\rightarrow ISM}}$ and thus the disc would be ejecting more gas than it would accrete. Because the mass in both the corona and the disc of the MW should be growing, $f_{\mathrm{cor}}$ is hence in the range $0\leq f_{\mathrm{cor}} \leq 1$ and represents the fraction of baryons accreted onto the halo over $\Delta t$ that end up in the corona. $f_{\mathrm{cor}}$ is related to the fraction of baryons accreted onto the halo that ends up in the disc (as either ISM or stars):
\begin{equation}
f_\mathrm{disc}\equiv \frac{\Delta M_{\mathrm{CGM\rightarrow ISM}} - \Delta M_{\mathrm{ISM\rightarrow CGM}}}{\Delta M_{\mathrm{IGM\rightarrow CGM}}}.
\end{equation}
If no gas is expelled from the halo, i.e. $\Delta M_{\mathrm{CGM\rightarrow IGM}}=0$, then $f_{\mathrm{cor}}=1-f_{\mathrm{disc}}$. Generally, $f_{\mathrm{cor}}=1-f_{\mathrm{disc}}-f_{\mathrm{out}}$ where
\begin{equation}
f_\mathrm{out}\equiv \frac{\Delta M_{\mathrm{CGM\rightarrow IGM}}}{\Delta M_{\mathrm{IGM\rightarrow CGM}}}.
\end{equation}
This is the fraction of gas accreted onto the halo that was later ejected from it.

$f_{\mathrm{disc}}$ is related to the \emph{instantaneous}  (i.e. in terms of immediate mass changes $\dot{M}$ rather than long time scale changes $\Delta M$) fraction of gas accreted onto the halo that reaches the disc $\zeta\geq f_\mathrm{disc}$. $\zeta$ is generally greater than $f_\mathrm{disc}$ because it only concerns infall onto the disc and so does not have the outflow term included in $f_\mathrm{disc}$. This parameter is an input for many analytical `bath-tub' equilibrium models of galaxy evolution \citep[e.g.][]{dave12,lilly13,dekel14}. It is generally found to be in the range $0.3-0.5$ for MW progenitors at $z<2$ in theoretic models and simulations \citep{dave12,lilly13,mitchell20}.

A standard assumption is that the gas infall from the IGM equals the DM accretion scaled by the cosmological baryon fraction $f_\mathrm{b}$. Hence, we may write eq. \eqref{eq:deltamcor} as
\begin{equation}
\label{eq:deltamcorfb}
\Delta M_{\mathrm{cor}}=f_\mathrm{cor} f_\mathrm{b}\Delta M_{\mathrm{DM}}.
\end{equation}
This is convenient since the growth of DM haloes is a robust prediction of cosmological simulations that agrees well between different simulations regardless of differences in their gas physics. Cosmological hydrodynamical zoom in simulations with AGN feedback suggest that eq. \eqref{eq:deltamcorfb} substantially overestimates the baryon infall at low halo masses because strong feedback is able to affect gas outside the virial radius and largely prevent it from being accreted onto the halo \citep{nelson15,hafen19,wright20}. However, the mass below which this effect becomes important depends on details in the feedback implementation and it is not clear if this effect is significant at the $M_{\mathrm{vir}} \gtrsim 10^{11.5} M_\odot$ virial masses at $z<2$ that we consider.


\bsp	
\label{lastpage}
\end{document}